\documentclass{aastex63}
\usepackage[all]{hypcap}
\usepackage{float}
\usepackage[T1]{fontenc}

\graphicspath{{./}{Figures/}}

\begin{document}

\title{Evolution of Elemental Abundances During B-Class Solar Flares: Soft X-ray Spectral Measurements with Chandrayaan-2 XSM}

\correspondingauthor{Biswajit Mondal}
\email{biswajitm@prl.res.in, biswajit70mondal94@gmail.com}

\author[0000-0002-7020-2826]{Biswajit Mondal}
\affiliation{Physical Research Laboratory, Navrangpura, Ahmedabad, Gujarat-380 009, India }
\affiliation{Indian Institute of Technology Gandhinagar, Palaj, Gandhinagar, Gujarat-382 355, India}
\author[0000-0002-4781-5798]{Aveek Sarkar}
\affiliation{Physical Research Laboratory, Navrangpura, Ahmedabad, Gujarat-380 009, India }
\author[0000-0002-2050-0913]{Santosh V. Vadawale}
\affiliation{Physical Research Laboratory, Navrangpura, Ahmedabad, Gujarat-380 009, India }
\author[0000-0003-3431-6110]{N. P. S. Mithun}
\affiliation{Physical Research Laboratory, Navrangpura, Ahmedabad, Gujarat-380 009, India }
\author[0000-0003-2504-2576]{P. Janardhan}
\affiliation{Physical Research Laboratory, Navrangpura, Ahmedabad, Gujarat-380 009, India }
\author[0000-0002-4125-0204]{Giulio Del Zanna}
\affiliation{DAMTP, Centre for Mathematical Sciences, University of Cambridge, Wilberforce Road, Cambridge CB3 0WA, UK}
\author[0000-0002-6418-7914]{Helen E. Mason}
\affiliation{DAMTP, Centre for Mathematical Sciences, University of Cambridge, Wilberforce Road, Cambridge CB3 0WA, UK}
\author[0000-0002-3064-0610]{Urmila Mitra-Kraev}
\affiliation{DAMTP, Centre for Mathematical Sciences, University of Cambridge, Wilberforce Road, Cambridge CB3 0WA, UK}
\author[0000-0001-9199-4925]{S. Narendranath}
\affiliation{Space Astronomy Group, U R Rao Satellite Centre, ISRO, Bengaluru-560 017, India}

\begin{abstract}

The Solar X-ray Spectrometer (XSM) payload onboard Chandrayaan-2 provides disk-integrated solar spectra in the 1 -- 15 keV energy range with an energy resolution of 180 eV (at 5.9 keV) and a cadence of 1~second. During the period from September 2019 to May 2020, covering the minimum of Solar Cycle 24, it observed nine B-class flares ranging from B1.3 to B4.5. Using time-resolved spectroscopic analysis during these flares, we examined the evolution of temperature, emission measure, and absolute elemental abundances of four elements -- Mg, Al, Si, and S.
These are the first measurements of absolute abundances during such small flares and this study offers a unique insight into the evolution of absolute abundances as the flares evolve. 
Our results demonstrate that the abundances of these four elements decrease towards their photospheric values during the peak phase of the flares. 
During the decay phase, the abundances are observed to quickly return to their pre-flare coronal values. 
The depletion of elemental abundances during the flares is consistent with the standard flare model, suggesting the injection of fresh material into coronal loops as a result of chromospheric evaporation. 
To explain the quick recovery of the so-called coronal ``First Ionization Potential (FIP) bias'' we propose two scenarios based on the Ponderomotive force model.
\end{abstract}


\section{Introduction}\label{sec-introduction}

The study of elemental abundances in different layers of the solar atmosphere is of great interest in solar physics. It can provide information about the mass and energy transfer between these layers. It has been observed that elements with a First Ionization Potential (FIP) less than 10 eV are 3-4 times more abundant, relative to those with high-FIP, in the quiescent cores of active regions, while their values in the quiet Sun are closer to their photospheric values.
This phenomenon is known as coronal FIP bias.
See, for example, \cite{Feldman_1992b,Feldman_2000,Feldman_2003,Saba_1995,delzanna:2013,Zanna_2014} and 
especially the recent review by \cite{Giulio_2018LRSP}. Most previous observations were however
unable to establish the absolute values of the abundances, as line-to-continuum 
measurements have been lacking. \cite{delzanna:2013} used instead an emission measure argument to 
show that low-FIP elements have  abundances  3.2 higher than their 
photospheric values in the 3 MK plasma of quiescent active region cores. Lower values, around 2, are 
found for the 2 MK plasma in the cores, while large abundance variations are found in other 
structures such as the 1 MK cooler loops or in brightenings. 

 Other stars also show chemical abundance variations, although generally active stars show an inverse FIP bias, i.e. the opposite of what is observed in quiescent active regions \citep{Drake_1994,Drake_1997,Laming_1999}.  Recently, the inverse FIP bias has also been
observed close to large Sunspots producing large flares \citep[see, e.g.][]{doschek_etal:2015}.
Several theories have attempted (e.g.,~\cite{hnoux_1998SSRv}) to explain the FIP bias, the most recent one being based on the Ponderomotive force model~\citep{Laming_2004,Laming_2009ApJ,Laming_2012ApJ, Laming2015,Laming_2017,Laming_2021ApJ},
which is able to reproduce the overall features of both the FIP and inverse FIP bias. 
Studies of chemical abundance variations are therefore important in general 
for astrophysics, as the observed variations are likely related to the (still unknown) physical processes that heat the plasma to coronal temperatures. 

By analyzing EUV low-temperature (less than 1 MK) spectral lines of flares and surges observed by Skylab, \cite{Feldman_90} \& \cite{Feldman_1992b} noted that the relative abundances of several elements (e.g., O and Mg) were close to their  photospheric values. These were interpreted as a consequences of chromospheric evaporation during the events, injecting photospheric plasma into the corona.

Later, with the improved spectroscopic capability in the X-rays and EUV, 
nearly photospheric abundances have been found for several large flares~\citep{delzanna_woods:2013,Warren_2014,Sylwester_2014, Sylwester_2015,Dennis_2015ApJ}. In these cases, the observations were of hot (about 10 MK) flare plasma, and 
line to continuum measurements indicated that indeed the absolute values 
were close to photospheric. We note, however, that a large scatter of values has been 
reported, partly due to the uncertainty in the continuum evaluation, partly due to the 
use of different analysis techniques.  
One interesting aspect of such studies is that none reported variations 
during the events, which are generally long-lasting for several hours.

Detailed studies of larger flares were possible because of their strong signal to noise ratio. However, it is also important to investigate their smaller counterparts. Observations show that smaller flares occur much more 
frequently than the bigger ones~\citep{Hudson_91}, so 
understanding their physics will be of great interest. As detailed observations 
have been lacking, it is not clear if the evolution of these smaller events follows 
the standard flare model.  It is only recently, with the advent of better instrumentation, that in-depth studies of smaller flares are becoming feasible~\citep{Kuhar_2018ApJ,Urmila_2019,Athiray_2020ApJ,cooper_2020ApJ,Glesener_2020ApJ,Duncan_2021ApJ,xsm_microflares_2021}. Lately, using X-ray time resolved spectroscopy,~\cite{Narendranath_2020} have carried out abundance studies of flares as small as GOES B9-class. Like the larger ones, these small flares have also shown a depletion of low FIP elements relative to H, during their evolution. 

The Solar X-ray Monitor (XSM) \citep{vadawale_2014,shanmugam_2020} onboard the Chandrayaan-2 provides soft X-ray disk-integrated solar spectra, with a good energy resolution (180 eV @ 5.9 keV) and high cadence (1 sec). Its high sensitivity and capability of measuring elemental abundances on an absolute scale (i.e., with respect to H) enable us to perform a comprehensive study of smaller flares.
The instrument has captured several sub-A class flares during the minimum of solar cycle 24~\citep{xsm_microflares_2021}. The statistical analysis of these tiny flares demonstrates the possible role of nanoflares in heating the quietest solar corona during this period. The instrument has also been used to study elemental abundances of the 2 MK quiescent solar corona during the aforementioned solar minimum~\citep{xsm_XBP_abundance_2021}. During this time, the coronal FIP bias was found to be about 2, less than the FIP bias of an active region (around 3-4). The overall emission of the global corona during this quiet phase of the Sun was found to be dominated by X-ray bright points.

 In this work, we present the evolution of temperature, emission measure, and absolute abundances of Mg, Al, Si, and S during a set of nine GOES B1.3-B4.5 class flares. These flares were observed during the minimum of solar cycle 24. At the time of these observations the Sun was extremely quiet, less dynamic and had single isolated active regions on the disk. For the first time, we show a clear and consistent variation of elemental abundances over the entire duration of these flares. 

The rest of the paper is organized as follows: 
 In Section~\ref{Observation} we present the observation, XSM data analysis, and identification of events. In Section~\ref{sec-TimeResolved_spectroscopy} we describe the details of the spectroscopic analysis.
After discussing the overall results in Section~\ref{results}, we summarize the article in Section~\ref{discussion}.

\section{Observations and Data Analysis}\label{Observation}

 XSM observes the Sun as a star from a lunar orbit. It  measures the soft X-ray spectrum in the 1 -- 15 keV energy range with an energy resolution better than 180 eV at 5.9 keV. 
 The unique design of XSM makes it possible to measure the X-ray spectrum with a
 stable energy resolution over a broad range of solar X-ray intensities, from lower than A-class up to X-class. 
For this purpose, it uses a Silicon Drift Detector (SDD) as the sensing device and a filter wheel mechanism along with the onboard logic for auto-detection of large flares to extend its dynamic range. In the present work, we focus on the largest observed flares, all of which are B-class. These flares were observed during the first two prime observing seasons of  the XSM from September 12 to November 20, 2019 (DD1) and February 14 to May 19, 2020 (DD2) -- see \cite{xsm_flight_performance} and \cite{vanitha20} for details on XSM observing seasons referred to as `dawn-dusk' (DD) and `noon-midnight' (NM) seasons. 

 XSM acquires spectra every second which are stored as raw, day-wise (level-1) data files. 
Due to the varying visibility of the Sun within the XSM Field Of View (FOV), there are periods when the Sun moves out of the FOV or is occulted by the Moon, and these periods provide the background observations. The raw data include both solar and background observations.
 The XSM Data Analysis Software (XSMDAS)~\citep{xsm_data_processing_2020} generates level-2 science data for durations for which the Sun was within the FOV, by selecting appropriate Good Time Intervals (GTI) based on the observing geometry and few other instrumental parameters. The default level-2 data contain the effective area corrected light curve of 1 second and 
spectra of 60 seconds cadence. XSMDAS also supports a user-defined GTI and time cadence to
 generate customized data products, which we utilize for the present analysis as discussed below.

\begin{figure*}[ht!]
\centering
\includegraphics[width=1.0\linewidth]{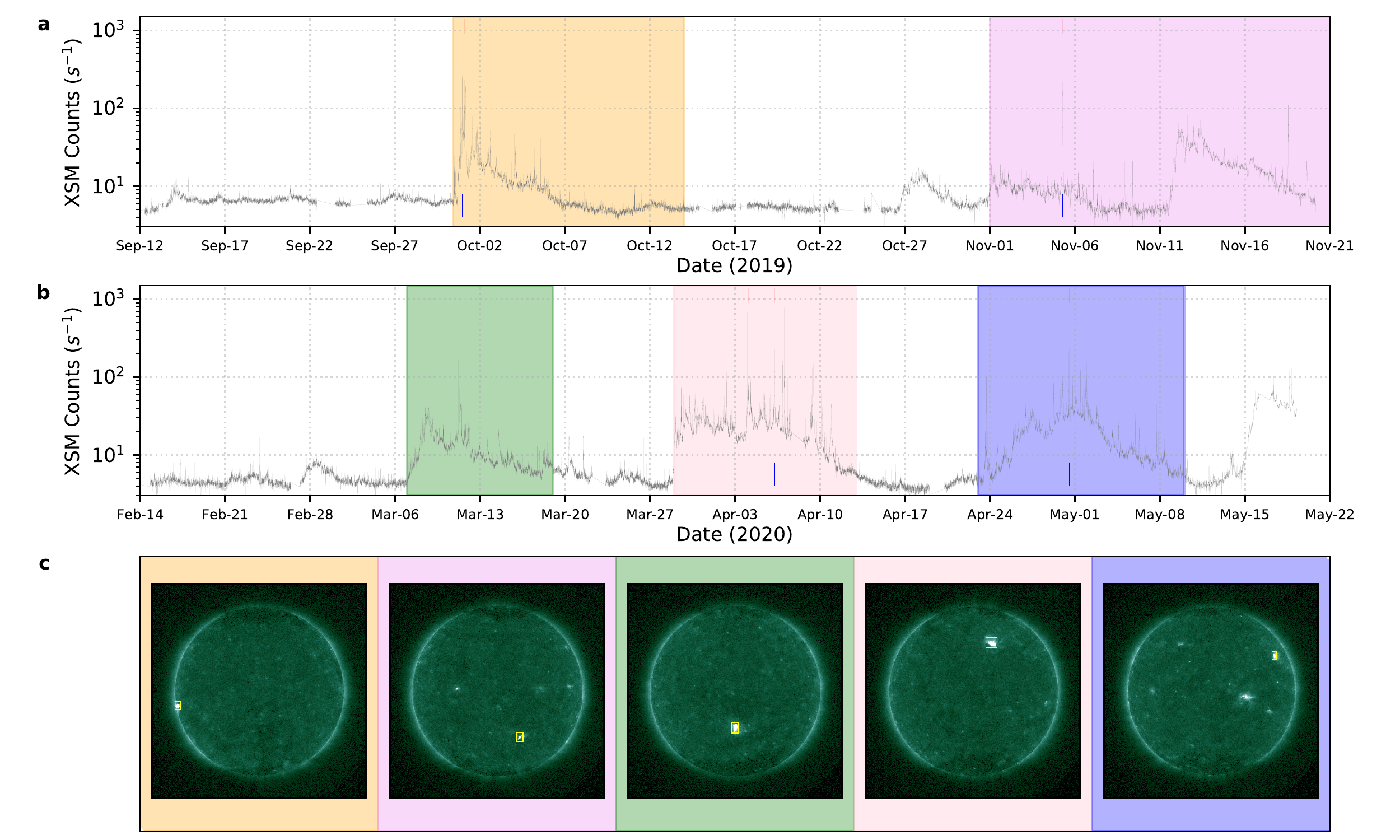}
\caption{Panels {\bf a} and {\bf b} show the X-ray light curve in the 1-15 keV energy range with a time cadence of 120~s, as 
measured by XSM during the two ‘Dawn-Dusk’ seasons (DD1 and DD2), respectively. 
Different color background shades represent periods when AR, recognized by National Oceanic and Atmospheric
 Administration (NOAA), are present on the solar-disk.
Panel {\bf c} shows representative full disk EUV images by AIA/SDO (94 \textup\AA\ channel) for the duration marked
 by the blue vertical line of the corresponding shaded regions of panel {\bf a} and {\bf b}.
Vertical red lines represent the peak time of the selected flares.
}
\label{fig-AR_LC}
\end{figure*}

Figure~\ref{fig-AR_LC} (panels {\bf a} and {\bf b}) shows the solar X-ray light curve observed by the XSM in the energy range 
of 1 to 15 keV during DD seasons DD1 and DD2. Shaded regions of the light curve show enhanced X-ray emission due to the
 presence of active regions (AR). The different colors represent different NOAA ARs, with each of them showing a number of
 flaring episodes. Since the objective of the present analysis is to investigate the temporal evolution of the flares, we
 select all the large flares for which the XSM count rate increases by more than 200 cps from the
 pre-flare baseline. We find a total of ten events satisfying this criterion and these events have been marked by vertical
 red lines in the figure. However, the flare of November 5, 2019, peaking at 06:11 UT was only partially observed by the XSM. We therefore exclude this event from further analysis. 

\begin{figure*}[ht!]
\centering
\includegraphics[width=1.\linewidth]{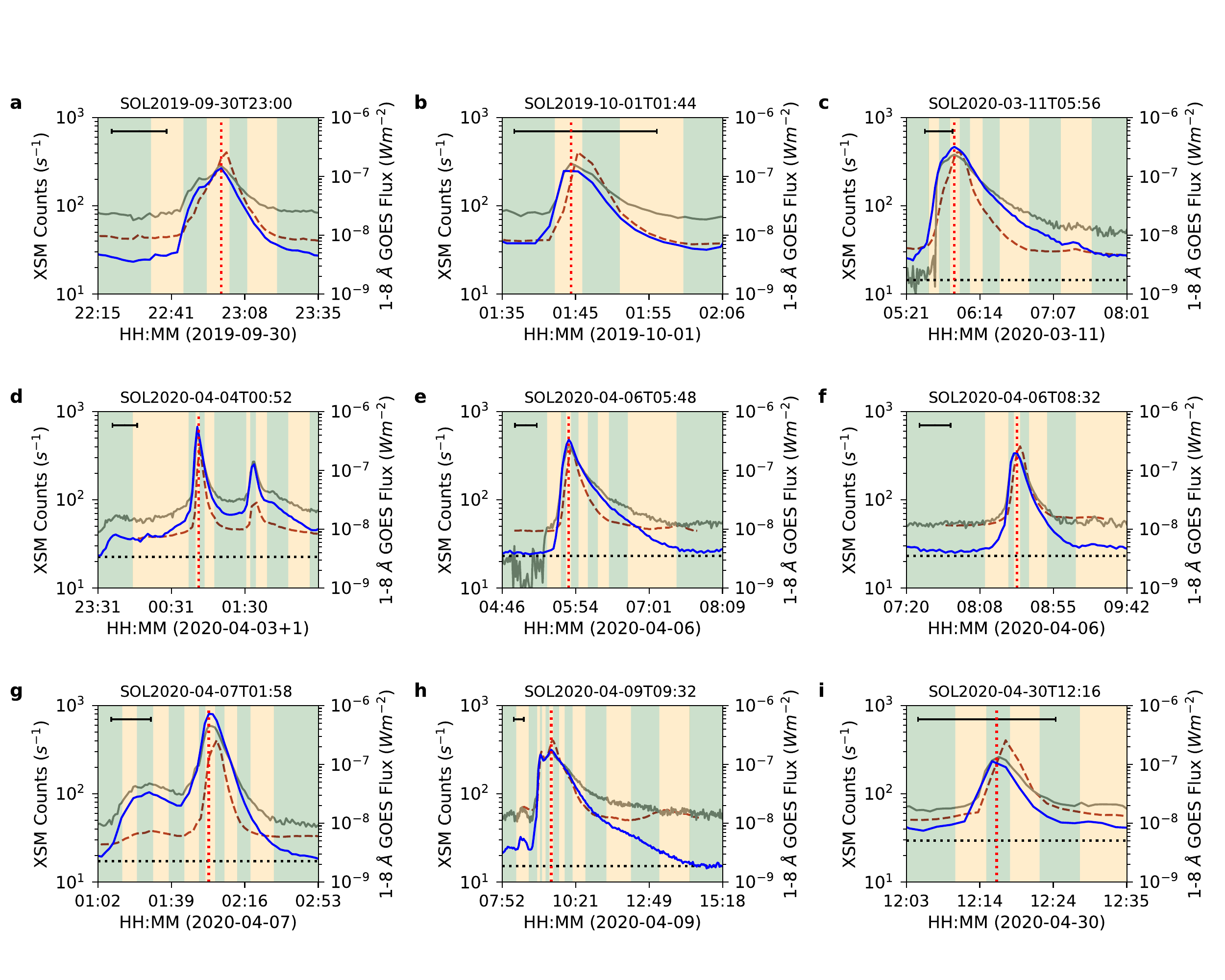}
	\caption{The 1-15 keV XSM X-ray light curves (blue) for the selected flares. The observed GOES-16 XRS fluxes of these 
flares are shown by grey lines while the brown dashed lines show the normalized (with XSM count rates) AIA 94~$\textup\AA$ fluxes, spatially integrated over the associated AR as marked by the yellow boxes in Figure~\ref{fig-AR_LC}c. 
Background green and orange shaded regions represent durations for which the integrated XSM spectra have been 
generated for time-resolved spectroscopy as described in Section~\ref{Observation}. The vertical red dotted lines
represent the flare peak time. Black horizontal dotted lines in panels {\bf c-i} are the average count rates for the non-flaring plasma
 as discussed in Section~\ref{sec-TimeResolved_spectroscopy}. The black scale-bars at the top left of each panel represent
 a 20-minute time interval on the x-axis. A flare ID, for each flare corresponding to the flare peak time in the format SOLyyyy-mm-ddThh:mm, 
is given on the top of every panel.}
\label{fig-BClassFlare_LC}
\end{figure*}

Figure~\ref{fig-BClassFlare_LC} shows an enlarged view of the X-ray light curves observed by the XSM (blue) for these 
nine flares, which have been designated with appropriate identifiers  
corresponding to the flare peak time, following the standard convention~\citep{Flare_identification}.
The figure also includes 1 -- 8 $\textup \AA$ GOES-16 X-Ray Sensors (XRS) light curves in grey color for comparison, showing that
 the nine selected flares are all small B-class events with the peak flux ranging from 1.34 $\times$ $10^{-7}$ W/m$^2$ to
 4.50 $\times$ $10^{-7}$ W/m$^2$.
It should be noted that some of the earlier studies (e.g. ~\citealp{Christe_2008,Hannah_2008}) have referred to such events
 as microflares; however, since the XSM has observed many weaker sub-A class events that have been referred to as microflares in an earlier paper 
 \citep{xsm_microflares_2021}, we continue to refer to the selected events as B-class flares.

Since the XSM observes disk-integrated X-ray emission from the Sun, we use Solar Dynamics Observatory/Atmospheric Imaging Assembly (SDO/AIA)~\citep{Lemen_2012SoPh} images in the 94~$\textup\AA$ channel to examine the spatial distribution of the enhanced activity on the solar disk.
AIA images at the time of the flares (examples in Figure~\ref{fig-AR_LC}c) show that only isolated ARs are present on the solar disk.
For each flare we derive the AIA 94 $\textup\AA$ light curve integrated over the regions around the respective AR (marked by small yellow boxes on the corresponding full disk AIA images in Figure~\ref{fig-AR_LC}c), they are shown with
 brown dashed curves in Figure~\ref{fig-BClassFlare_LC}. 
 Their similarity with the XSM light curves confirm that the
 entire enhancement of X-ray intensity observed by the XSM during these flares originated within the respective AR. This
 confirms that the disk-integrated XSM observations can be used to study the temporal evolution and properties of the
 flaring plasma. 

\begin{figure*}[ht!]
\centering
\includegraphics[width=1.\linewidth]{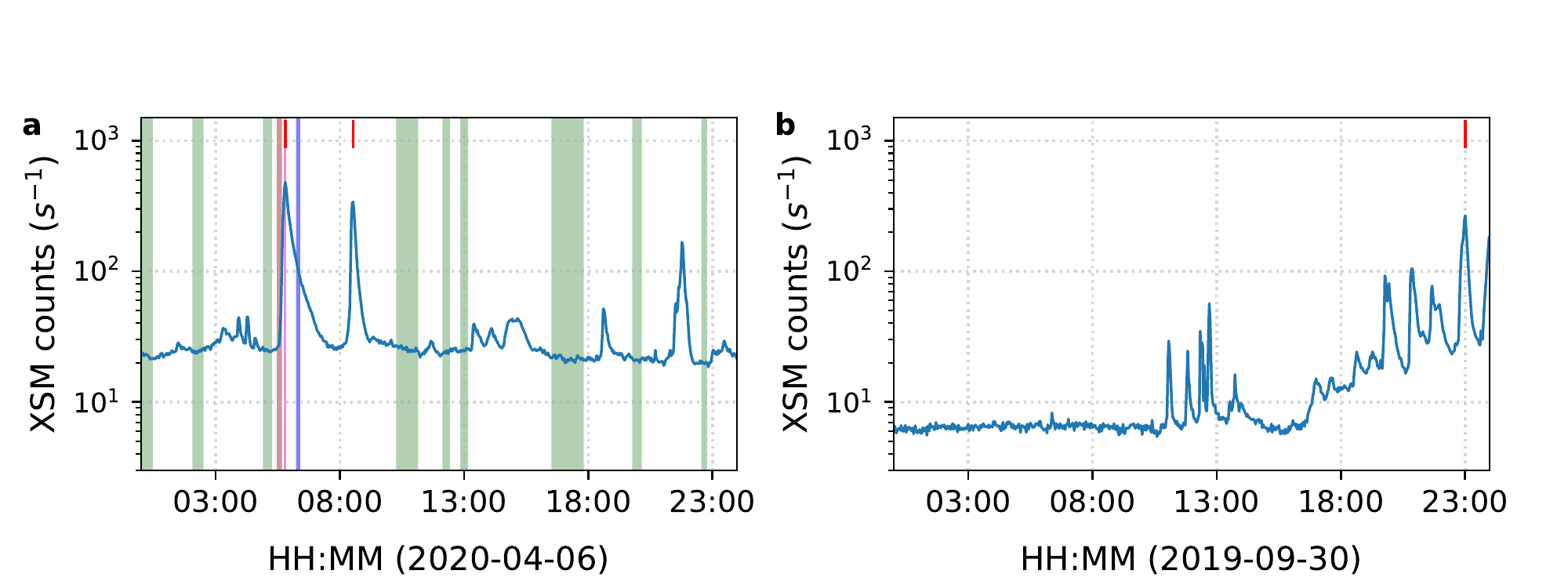}
\caption{Full day XSM 1-15 keV solar X-ray light curve on 2020-04-06 ({\bf a}) and 2019-09-30 ({\bf b}). Selected flare
 peak times for the same day are shown by the red vertical lines. In panel {\bf a} the green shaded regions demarcate the
 time durations used to generate the non-flaring emission spectrum required
 for the two-temperature spectral fitting for the flares SOL2020-04-06T05:48 and SOL2020-04-06T08:32. 
The brown, violet, and blue shaded regions show the time interval of the spectra shown in Figure~\ref{fig-Fit_2T_20200406054800}.
Panel {\bf b} shows multiple small flaring episodes before and after the flare SOL2019-09-30T23:00.
All of these flares occur within the same AR, which appeared at the eastern limb of the Sun.
Because of this, a reliable non-flaring emission could not be established for this flare. 
}
\label{fig-BKG_emission_GTI}
\end{figure*}

In order to investigate the temporal evolution during the selected flares, we divide each flare duration into multiple smaller
 time intervals corresponding to various phases of each flares. 
The intervals for each flare, shown as alternate background shades in Figure~\ref{fig-BClassFlare_LC}, are selected based
 on visual inspection of the XSM light curve in the energy range of 1 -- 15 keV, such that all intervals have a sufficient
 number of counts to perform spectral analysis.
These time intervals are then used as user GTI in XSMDAS to generate a series of spectra for all nine flares. The modelling
 of these spectra is then carried out using two different approaches, as discussed in detail in the next section. The first
 approach assumes that the emission arises from plasma having two distinct components corresponding to the flaring
 plasma and the rest of the coronal plasma, respectively. The second approach assumes that the emission arises
 from a single component isothermal plasma. 
The first approach requires spectrum corresponding to a non-flaring plasma, which is generated for the non-flaring time intervals as shown in Figure~\ref{fig-BKG_emission_GTI}a for two representative flares, viz. SOL2020-04-06T05:48 and SOL2020-04-06T08:32. The blue line shows the full day light curve while the green shaded regions show periods for which the non-flaring emission spectrum is generated. 
Such non-flaring time intervals are visually selected from the corresponding daily light curves for all flares, with the
 exception of the two flares SOL2019-09-30T23:00 and SOL2019-10-01T01:44. For these two flares, it is difficult to select
 a reliable non-flaring duration because of the  multiple small flaring episodes occurring within the same AR as shown  in
 Figure~\ref{fig-BKG_emission_GTI}b. Moreover, the AR starts appearing from the eastern limb of the sun during
 these two events. Hence, the spectral analysis of these flares is restricted to the isothermal plasma assumption (second approach).  

\section{Time resolved Spectral Analysis}\label{sec-TimeResolved_spectroscopy} 

Soft X-ray spectra of the solar corona provides information on the physical conditions 
and radiative processes in the emitting plasma. It typically consists of a continuum
arising from free-free, free-bound, and two-photon radiative processes superimposed by 
emission lines corresponding to different ionization states of various elements. 
The expected spectrum for a given set of physical parameters such as temperature (T), 
density (usually represented as emission measure, EM) as well as the abundances of various 
elements, can be analytically calculated \citep{Giulio_2018LRSP}. This analytical or 
synthetic spectrum, when fitted to the observed spectrum, can constrain the physical 
parameters of the observed plasma. We use the CHIANTI database  
version~10~\citep{chiantiV10_Zanna2020} to generate the synthetic spectra and 
XSPEC (an X-ray spectral fitting package~\citep{ref-xspec}) to fit the spectrum. 
Since the synthetic spectrum calculated by CHIANTI can not be directly used in XSPEC for
spectral fitting, we have developed a local XSPEC model, {\em chisoth}, from a wide range of 
pre-calculated spectra, as described in Appendix~\ref{sec-model} 
(the same model was used for the spectral analysis of the quiet Sun and microflare 
spectra presented by \cite{xsm_microflares_2021,xsm_XBP_abundance_2021}). 

The main objective of our analysis is to study the evolution of plasma properties during
the course of the nine selected B-class flares. Since these are relatively small flares,
it is likely that during different flare phases there are considerable contributions 
from non-flaring plasma. Usually, such contributions are subtracted from the flare spectrum by considering a pre-flare spectrum as background. 
However, this is not preferable in the case of XSM because the sun angle, and thus the effective area of the instrument, changes continuously. 
Hence, we model the spectrum of the non-flaring plasma independently and include it as a fixed component while fitting the flare spectrum. An additional advantage of this approach is that in most cases it is possible to co-add several epochs having the same intensity and spectra, as shown in Figure~\ref{fig-BKG_emission_GTI}a, to increase the statistical significance of the non-flaring spectrum. 
We then fit this non-flaring X-ray spectrum with our isothermal model. 
Though the model allows one to vary the abundances of all elements in the range of Z=2 to 30, we retain the abundances of only Mg, Al, and Si, for which the line features are observable in the spectra, as free parameters.
Abundances of the remaining elements are fixed to their respective coronal values taken from the combined dataset ``sun\_coronal\_1999\_fludra\_ext.abund" (hereafter A\_F99) available within the CHIANTI package.
These are so-called `hybrid' abundances, where the abundances of low (high) FIP elements were increased (decreased) compared to their photospheric values, by about a factor of two. 
The other two free parameters of the model are temperature (T) and emission measure (EM). We carry out the spectral analysis to determine the T, EM and abundances of Mg, Al, and Si of non-flaring plasma for all the flares except SOL2019-09-30T23:00 and SOL2019-10-01T01:44 as discussed in Section~\ref{Observation}. Table~\ref{table-I} shows the best fitted parameters for all the non-flaring spectra.

\begin{figure*}[ht!]
\centering
\includegraphics[width=1.\linewidth]{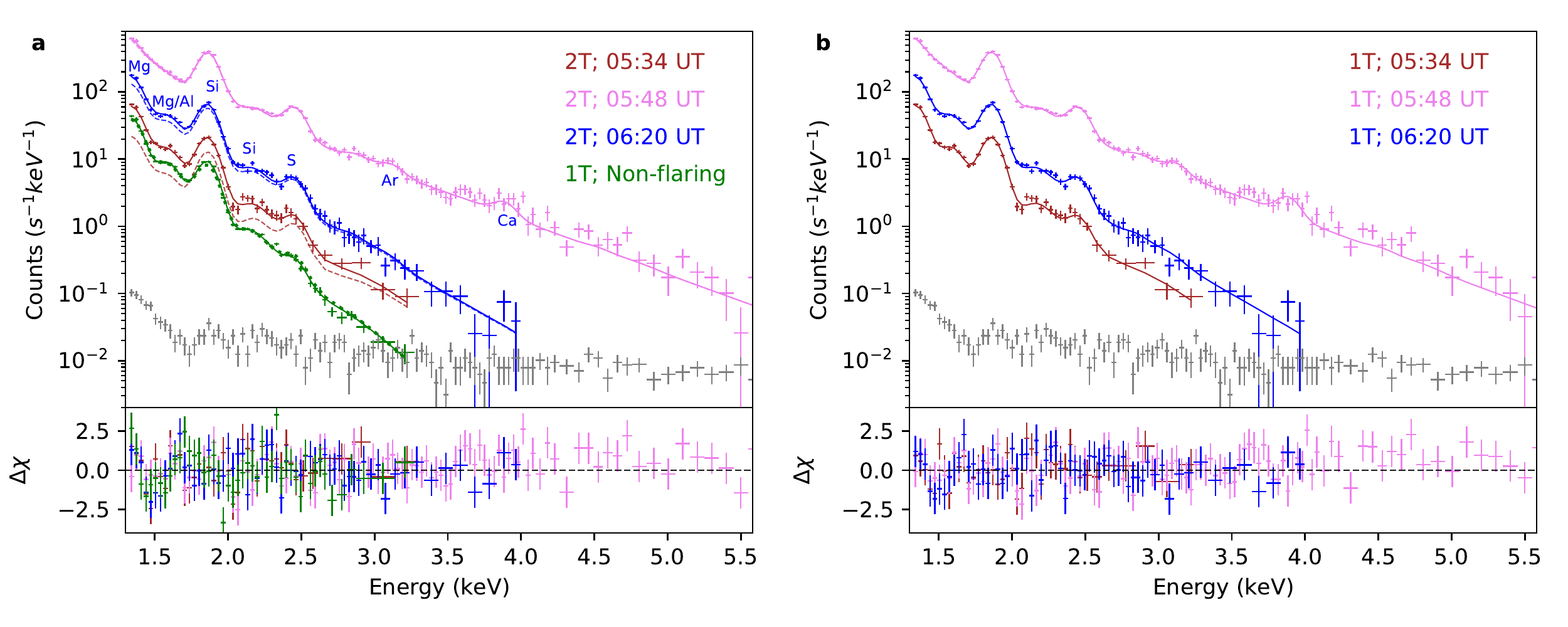}
\caption{Fitted  spectrum for the Flare SOL2020-04-06T05:48. The brown color represents  the 
rising phase at 05:34 UT, violet is for the flare peak at 05:48 UT and blue is for the decay 
phase at 06:20 UT. Panel {\bf a} shows the fitted spectra using a two temperature  isothermal 
emission model (solid lines), while  panel {\bf~b} shows the fitted spectra using a single 
temperature model (solid lines).
The green color in panel {\bf a} represents the fitted spectrum for the emission due to non-flaring plasma considered in the two component model. The dashed lines in panel {\bf a} represent modeled spectra of the flaring emission component.
The grey points show the non-solar background spectrum acquired for the times when the Sun was out of the XSM FOV. The energy ranges of the respective spectral fits were restricted to the energies where the solar X-ray spectrum dominates over the non-solar background.}

\label{fig-Fit_2T_20200406054800}
\end{figure*}

Next, we focus on the spectra generated for the different phases of the selected nine flares as 
shown in Figure~\ref{fig-BClassFlare_LC}. We fit each spectrum with a model consisting of two isothermal components. All parameters of the first component are fixed to their respective values obtained from non-flaring intervals, whereas for the second component, T, EM, Mg, Al, and Si are kept as free parameters and fitted for each spectrum. We also keep S abundance as a free parameter whenever fitting for S is statistically feasible. 
During the flare peak, occasionally Ar and Ca line complexes are also visible,  however, because of their  poor statistics, including them in the spectral fits and deriving their abundances leads to large error bars. Hence, we fix their abundances as a constant parameter during spectral fitting.  We have verified that keeping abundances of these two elements either as free or fixed do not affect any other fit parameters.
The one sigma uncertainties on all free parameters are obtained using the standard procedure in XSPEC. The representative spectral fits for the flare  SOL2020-04-06T05:48, for the three time durations (marked by brown, violet, and blue shaded regions in Figure~\ref{fig-BKG_emission_GTI}a) are shown in Figure~\ref{fig-Fit_2T_20200406054800}a. The fit results for all phases of the same flare are shown in Figure~\ref{fig-fitpar_20200406054800}.

For the two flares SOL2019-09-30T23:00 and SOL2019-10-01T01:44, it was not possible to 
carry out the spectral analysis with the two-component model due to the lack of reliable non-flaring spectra. Hence, for these two flares, we resorted to using only a model with a single temperature. However, in order to investigate any differences between the results obtained by the two spectral fitting approaches, we also carried out a time resolved spectral analysis, for all nine flares, under the isothermal assumption with a single component model.

The spectral fits with the single component model for the same spectra of the flare 
SOL2020-04-06T05:48 are shown in
 Figure~\ref{fig-Fit_2T_20200406054800}b and the fit results for all phases of the flare are 
over-plotted in Figure~\ref{fig-fitpar_20200406054800}. 
We find that, for all flares where both the two component and single component analysis is possible, the results obtained
 with both approaches are similar within error-bars.
Table~\ref{table-I} shows the values for the time intervals containing the  
temperature and emission measure peaks.  
This suggests that even for these small B-class flares, the emission from the flaring plasma 
dominates over the emission from
the non-flaring plasma and hence for the flares SOL2019-09-30T23:00 and SOL2019-10-01T01:44, 
the results of the single component analysis also represents variations 
associated to the flare loops.

\section{Results and Discussion}\label{results}

\begin{figure}
\begin{center}
\includegraphics[scale=0.7]{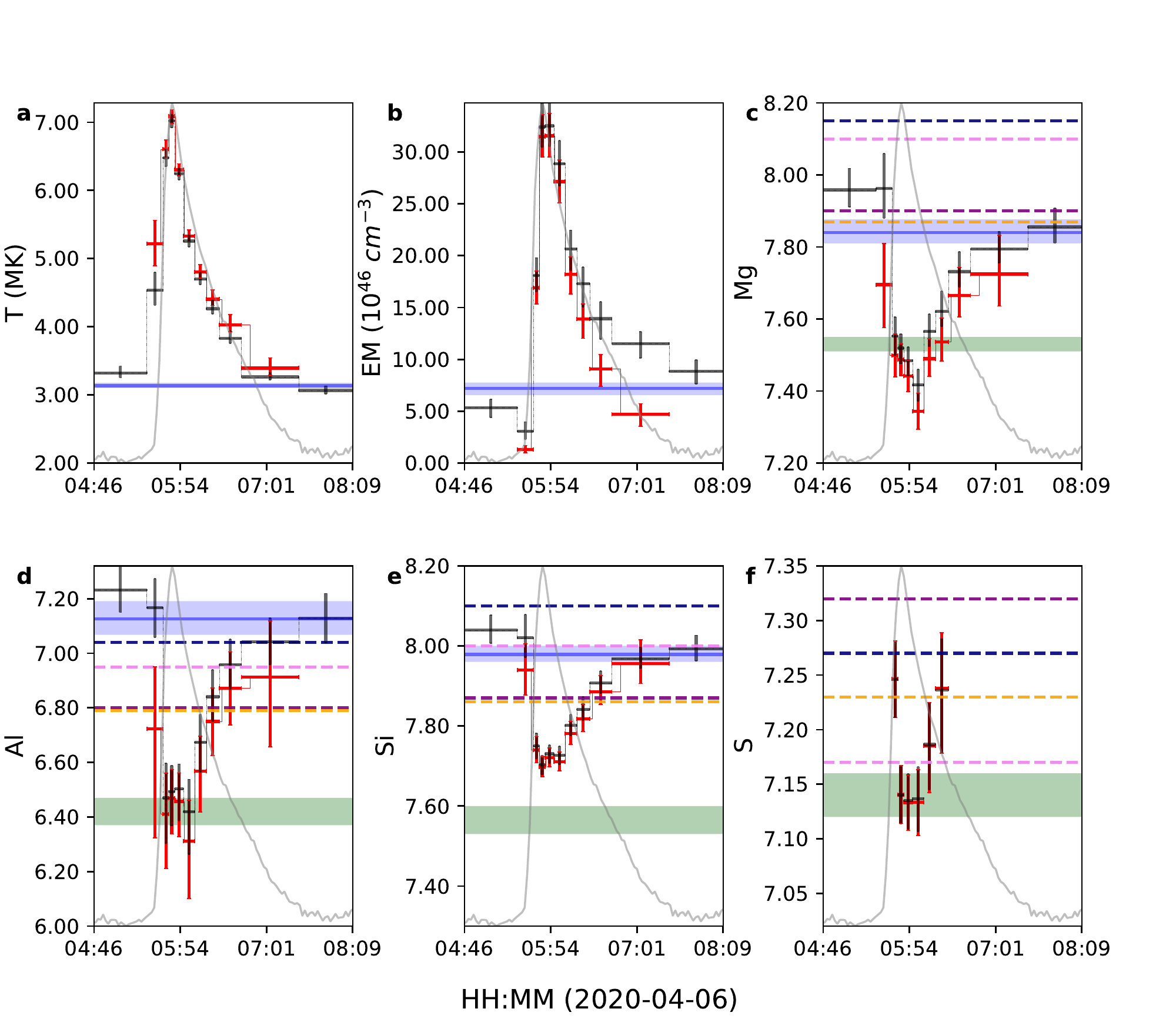}
\caption{The six panels show the results of the time resolved X-ray spectroscopy for the flare SOL2020-04-06T05:48. Panels {\bf a-b} show the variation of temperature and emission measure, respectively, during flare activity, while panels {\bf c-f} show the variation of elemental abundances of Mg~({\bf c}), Al~({\bf d}), Si~({\bf e}), S~({\bf f}) in logarithmic scale as described in Equation~\ref{abund_eq}.
Spectra are fitted by both a single component isothermal model and a two component isothermal model, as discussed in the text. 
The black and red points represent the best fit parameters obtained from the fitted spectra using a one component model (black) and a two component model (red), respectively.
The y-error bars represents 1$\sigma$ uncertainties for each of the parameters, whereas the x-error bars represent the duration over which a spectrum is integrated.
For a quick comparison with the reported abundance values for these elements, the corresponding panels ({\bf c-f}) also show the
coronal value of abundances compiled in the CHIANTI database of ``sun\_coronal\_feldman\_1992\_ext.abund" (A\_F92, navy blue), ``sun\_coronal\_fludra\_1999\_ext.abund" (A\_F99, Purple), and ``sun\_coronal\_schmelz\_2012.abund" (A\_S12, orange) and 
also reported by  \cite{delzanna:2013} (violet).
The range of reported photospheric abundances from various sources compiled in the CHIANTI database are shown as green bands.
The X-ray XSM light curve is over plotted in grey color in the background.\label{fig-fitpar_20200406054800}}

\end{center}
\end{figure}

\begin{figure}
\begin{center}
\includegraphics[scale=0.6]{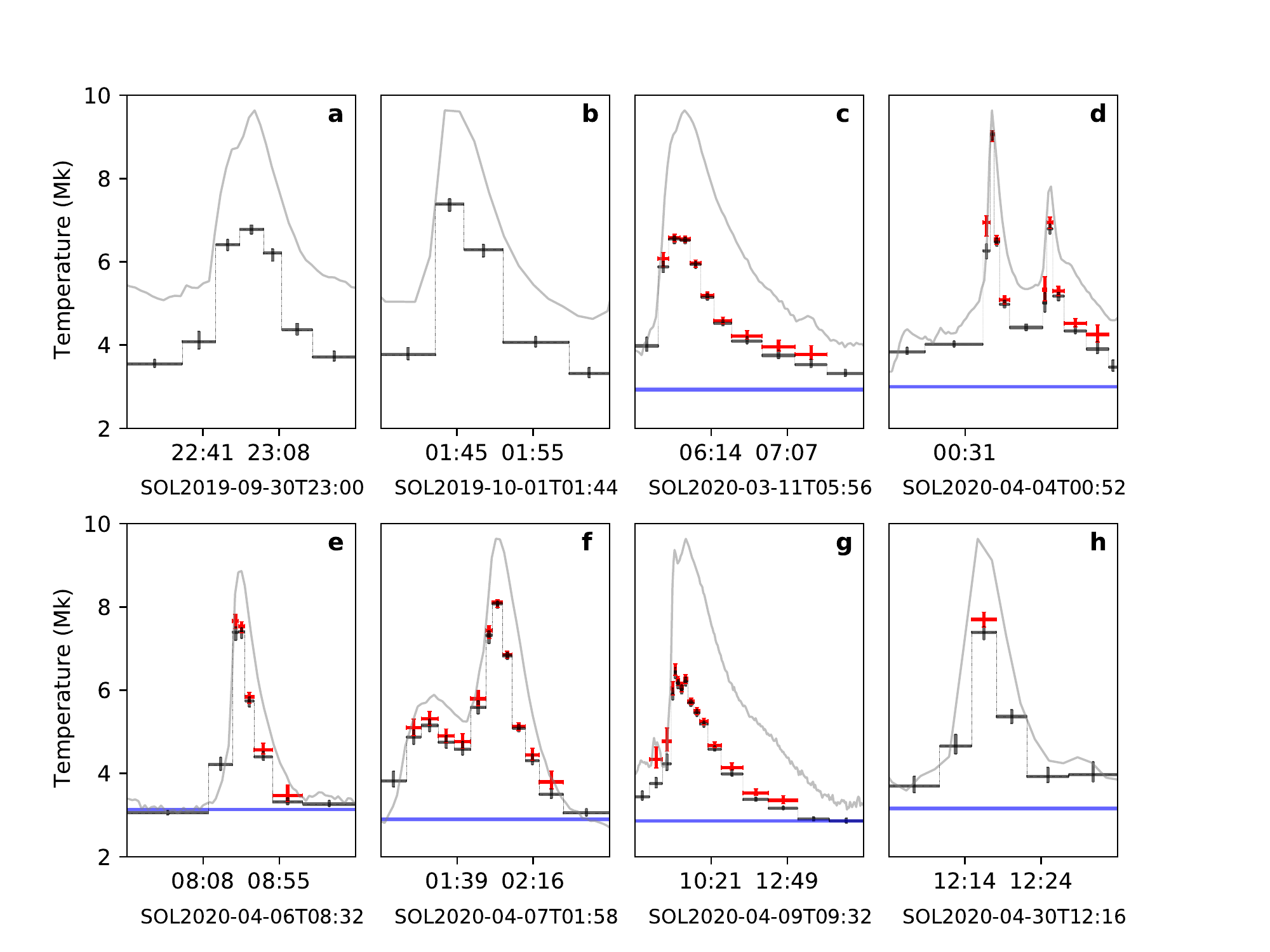}
\caption{Temperature evolution similar to Figure~\ref{fig-fitpar_20200406054800}a, but for different flares.\label{fig-fitpar_AllT}}
\end{center}
\end{figure}

\begin{figure}
\begin{center}
\includegraphics[scale=0.6]{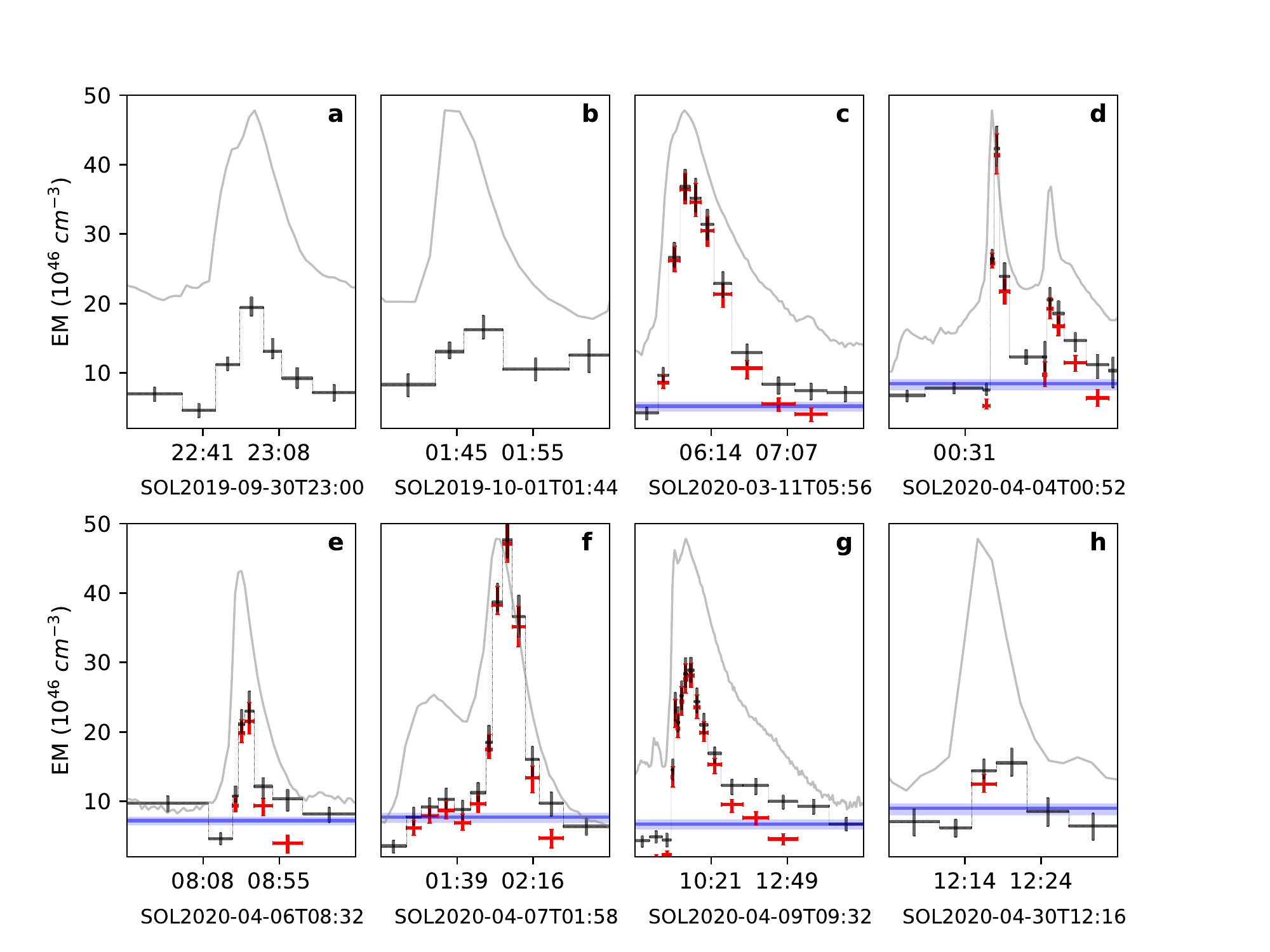}
\caption{Evolution of the emission measure similar to Figure~\ref{fig-fitpar_20200406054800}b, but for different flares.\label{fig-fitpar_AllEM}}
\end{center}
\end{figure}

\begin{figure}
\begin{center}
\includegraphics[scale=0.6]{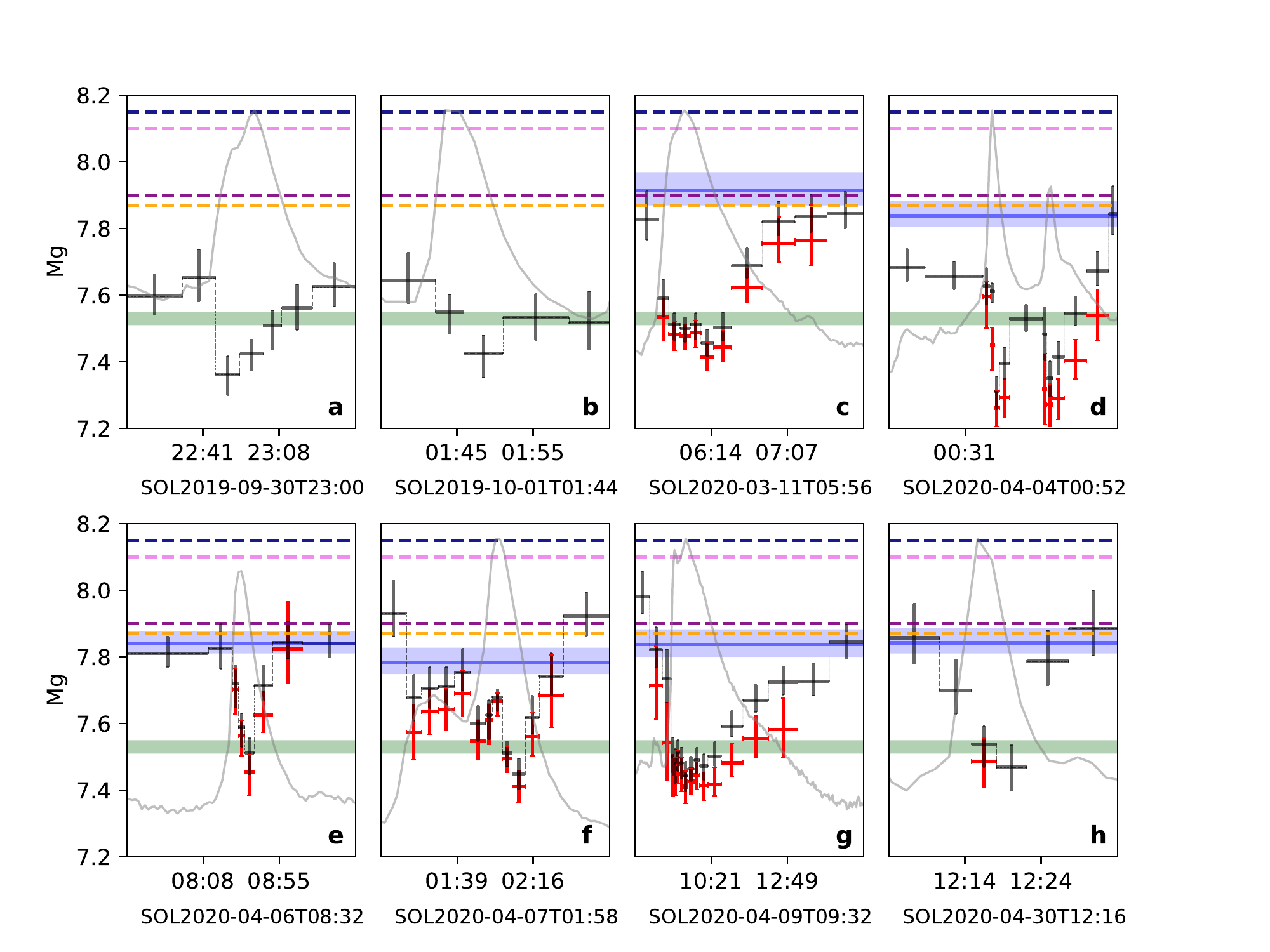}
\caption{Evolution of the absolute Mg abundance similar to Figure~\ref{fig-fitpar_20200406054800}c, but for different flares.\label{fig-fitpar_AllMg}}
\end{center}
\end{figure}

\begin{figure}
\begin{center}
\includegraphics[scale=0.6]{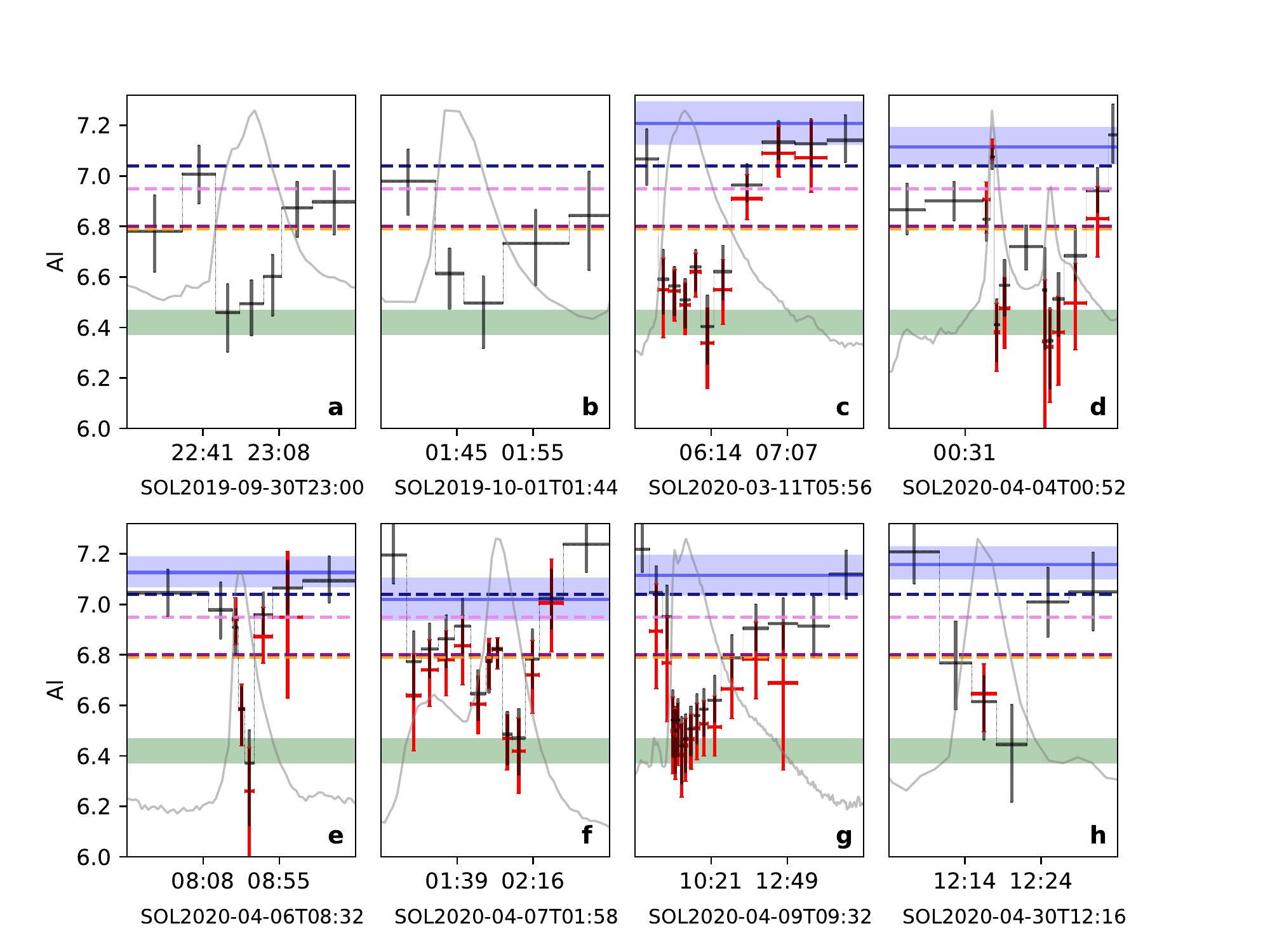}
\caption{Evolution of the absolute Al abundance similar to Figure~\ref{fig-fitpar_20200406054800}d, but for different flares.\label{fig-fitpar_AllAl}}
\end{center}
\end{figure}

\begin{figure}
\begin{center}
\includegraphics[scale=0.6]{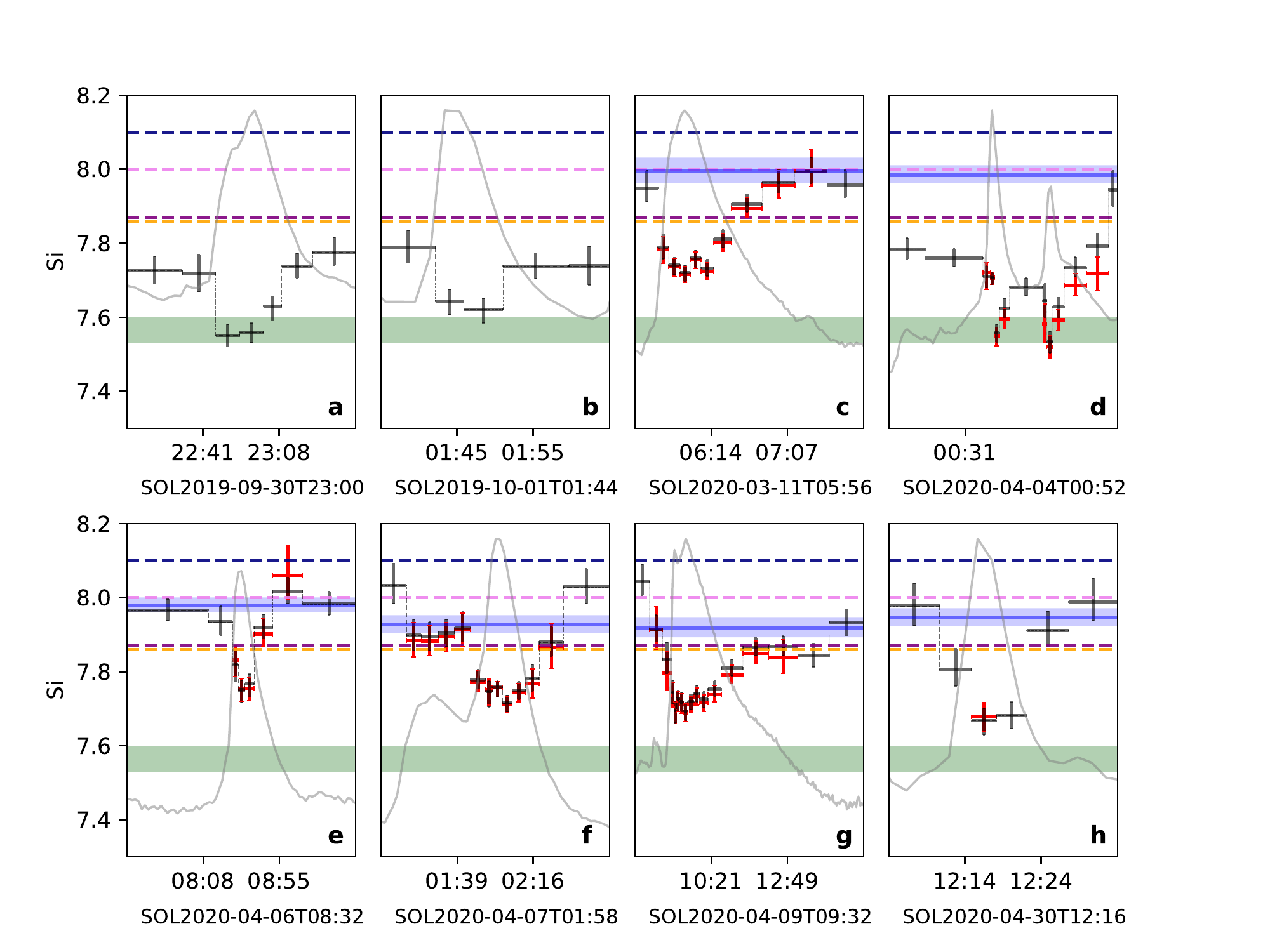}
\caption{Evolution of the absolute Si abundance similar to Figure~\ref{fig-fitpar_20200406054800}e, but for different flares.\label{fig-fitpar_AllSi}}
\end{center}
\end{figure}

In this study, we have performed a spectroscopic analysis of nine B-class flares. 
Using time-resolved spectroscopy of these flares, we have shown the
evolution of elemental abundances, temperature, and emission measure over
their lifetime.
Results from a representative flare (SOL2020-04-06T05:48) are demonstrated in Figure~\ref{fig-fitpar_20200406054800}.
The different panels demonstrate the time evolution of temperature (a), emission measure (b), and abundances of four elements (c-f), namely, Mg, Al, Si \& S. Temperature (Figure~\ref{fig-fitpar_AllT}), emission measure (Figure~\ref{fig-fitpar_AllEM}), and abundance (Figure~\ref{fig-fitpar_AllMg}-\ref{fig-fitpar_AllSi}) evolution of all other flares are shown separately. In the background of each plot, the light curve of the respective flare is shown with  a solid grey line. 

Temperatures (Figures~\ref{fig-fitpar_20200406054800}a, Figure~\ref{fig-fitpar_AllT}) and emission measures (Figures~\ref{fig-fitpar_20200406054800}b, Figure~\ref{fig-fitpar_AllEM}) of these flares rise with their soft X-ray activity, showing maxima near the time of the flare peaks.
In most of the flares, the temperature peaks lie ahead of the emission measure peaks. These can be identified by comparing Figure~\ref{fig-fitpar_AllT} with Figure~\ref{fig-fitpar_AllEM} (or Table~\ref{table-I}). However, two flares, namely, SOL2019-09-30T23:00 and SOL2020-03-11T05:56 (panels a and c) do not follow the trend, possibly due to larger time bins at the flare peak compared to the small flare duration. 
During the impulsive phases, elemental abundances (Figure~\ref{fig-fitpar_20200406054800}(c-f), Figures~\ref{fig-fitpar_AllMg}-\ref{fig-fitpar_AllSi}) are seen to reduce quickly from the coronal to near photospheric values, while the minima occurring almost simultaneously with the emission measure peaks. However, the abundances swiftly return to the coronal values during the decay phases of the flares.

Considering our `Sun as a star' observation, one may think the quick recovery of the coronal abundances may not be physical but merely be an effect of emission swapping from the flaring loop to the rest of the corona. However, results derived using our two temperature models, where emission from flaring and non-flaring plasmas are treated separately, show a similar abundance recovery (red points in Figure~\ref{fig-fitpar_20200406054800}(c-f), Figures~\ref{fig-fitpar_AllMg}-\ref{fig-fitpar_AllSi}). This confirms that this abundance recovery is not due to emission swapping.

Following the standard flare model (CSHKP:~\cite{Carmichael_1964,Sturrock_Sturrock1966_Nature,Hirayama_1974,Kopp_1976}), the evolution of temperature and emission measure can be explained by invoking chromospheric evaporation. After the flare onset, both temperature and emission measure start rising.  
When the coronal temperature achieves its peak, the heat energy from the reconnection site travels down to the chromosphere and starts evaporating chromospheric plasma into the loop with high velocity~\citep{Antonucci_1985}.
The peak of the emission measure is recorded once the loop gets filled with the high-density heated evaporated material~\citep{Ryan_2012,klimchuk_2017}. As the evaporative upflow can bring in fresh unfractionated chromospheric material into the corona, it can also explain the quick abundance depletion. However, explanation of the fast recovery (in minutes time scale) of the coronal FIP bias during the decay phases of the flares demands the support of additional mechanisms. 

\begin{figure}
\begin{center}
\includegraphics[scale=0.6]{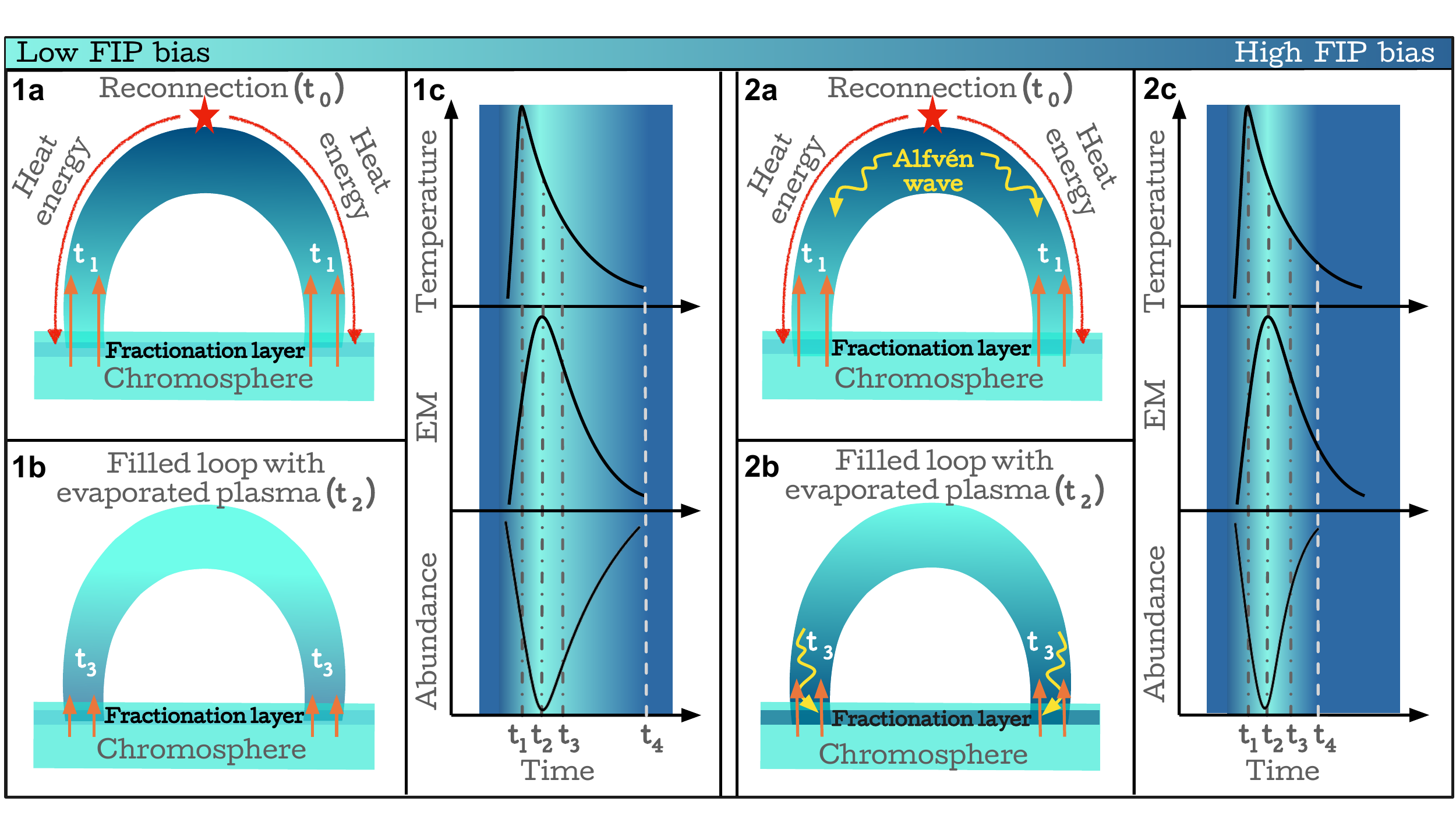}
\caption{{Schematic representation of flaring loop dynamics with lighter colors representing low FIP bias, and darker colors indicating high FIP bias. \textbf{Scenario 1:} (1a) Impulsive phase -- After the flare onset (at time $t_0$), heat energy travels down to the chromosphere and evaporates the plasma there.  The temperature peak is identified at the start (time $t_1$) of the chromospheric evaporation. During the initial phase, plasma evaporates with high velocity and quickly passes through the fractionation layer without getting fractionated, resulting in depletion of the coronal FIP bias. The Emission Measure (EM) peak is achieved (time $t_2$) once the loop gets filled with the chromospheric plasma. At around the same time, the abundances of the FIP elements also attain their minima. (1b) Decay phase -- The speed of the evaporative upflow slows down significantly allowing plasma enough time in the fractionation layer to get fractionated. The fractionated plasma eventually fills the coronal part of the loop thereby demonstrating the recovery of the coronal FIP bias. (1c) Representative curves of temperature, Emission Measure (EM), and elemental abundance evolution. Time $t_4$ marks the time of complete recovery of the coronal FIP bias. 
\textbf{Scenario 2:} (2a) At the time of the flare onset (time $t_0$), high amplitude Alfv\'en waves are initiated. The released heat energy gets transported to the chromosphere by suprathermal electrons, at speeds faster than the Alfv\'en waves. Evaporation starts at time $t_1$, and reduces the coronal FIP bias. The temperature peak is observed around this time. (2b) The EM peak and abundance minima are achieved when the loop is filled with the evaporated plasma ($t_2$). Once the flare-driven Alfv\'en waves arrive at the chromosphere (time $t_3$), they rapidly fractionate the plasma. These Alfv\'en waves also carry significant heat from the flaring site to assist the evaporation. The fractionated plasma is then transported to the coronal part of the loop through evaporative upflows, which help in rapidly recovering the coronal FIP bias. (2c) Representative curves of Temperature, EM, and elemental abundances for the second scenario.}
\label{fig-flaremodel}}
\end{center}
\end{figure}

One of the most widely accepted theories to explain the FIP bias in the closed loop coronal plasma is based on the Ponderomotive force model. The said force originates at the chromospheric layers of the magnetic flux tubes due to multiple reflections of the coronal Alfv\'en waves~\citep{Laming_2004}. According to this theory, the  FIP dependent elemental fractionation arises because the upwards Ponderomotive forces act only on the ions. Since the density and temperature of the chromospheric layer favor the ionization of low FIP elements, only these elements get fractionated there. Low FIP ions then experience the Ponderomotive force and drift upwards, enhancing their concentration in closed coronal loops. Observations suggest it takes several days for the FIP bias to get established in non-flaring active regions~\citep{Widing_2001}, although we note that the Skylab measurements discussed by ~\cite{Widing_2001} refer to cool transition region lines and the results are not directly related to the abundances of the hot core loops. In this context, two possibilities can be thought to be responsible for the quick transitions (order of minutes) of the coronal FIP bias during the impulsive and decay phases of all observed flares in this study. The schematic representations of these two possibilities are presented in Figure~\ref{fig-flaremodel}.

The first possibility envisages that the newly evaporated chromospheric material comes from a region beneath the fractionation layer, having almost non-fractionated photospheric plasma. It has been observed earlier that the injection of fresh material continues throughout the duration of the flare~\citep{Zarro_1988,Czaykowska_1999}. The injection velocity is highest 
during the impulsive phase of the flare and gradually reduces towards the decay phase~\citep{Fletcher_2011}. At the time of the initial phase, when the upflow velocity is high, plasma does not get enough time to spend in the fractionation layer to get fractionated and thus unfractionated photospheric plasma is injected 
into the coronal loop -- exhibiting a depletion of the coronal FIP bias. Once the flow speed reduces below $\sim 1$ km/sec~\citep{Laming_2004,Laming_2017}  at the top of the chromosphere, it is plausible to assume that the evaporated material takes longer to cross the thin fractionation layer and gets fractionated while traveling through it. In this scenario, the quick enhancement of the FIP bias is not only due to the slow drift of the ions but also due to the bulk velocity of the already (partly) fractionated material filling the loop. As the velocity of the evaporated material decreases, the plasma gets fractionated to a higher degree, and thus the FIP bias is restored to the pre-flare level by the time the flare emission decays completely.

The second possibility relies on the flare driven high amplitude Alfv\'en waves. Immediately after the flare, heat energy carried by the suprathermal electrons~\citep{Benz_2017} may travel faster towards the loop footpoints than the Alfv\'en waves to evaporate plasma from the chromosphere. The evaporated chromospheric non-fractionated plasma is responsible for the initial depletion of the FIP bias. However, once the newly generated Alfv\'en waves reach the fractionation layer, they start fractionating plasma through the generation of the Ponderomotive force. The high amplitude nature of these waves help to  fractionate plasma in a faster time scale~\citep{Dahlburg_2016}. These Alfv\'en waves may also carry a significant fraction of the released magnetic energy from the flaring site to the chromosphere, which may also assist the evaporation~\citep{Fletcher_2008}. The fractionated plasma then gets transported to the coronal loop, assisted by the evaporative upflow, which helps to recover the coronal FIP bias in a faster time scale.

Although all the observed flares show a similar variation of the abundance evolution, the abundance recovery time scale during the decay phase varies from flare to flare. In general, it ranges between $\sim$10 to $\sim$90 minutes. The second scenario, which explains the faster fractionation as well as transport, appears to be a more plausible mechanism to explain the faster recovery of the FIP bias. Whereas, if the recovery happens over a longer time scale ($\sim$ several hours) it can be explained through the former. Reconnection driven Alfv\'en waves are yet to be observed in flaring loops. However, the observations presented in this article imply their role in altering the coronal FIP bias.
XSM observes the Sun as a star, therefore misses any spatial information. On the other hand, it is to be noted from SDO observations that the flaring reconnection and the post-flare loops have fairly complicated geometric structures. Though the schematics of Figure~\ref{fig-flaremodel} consider a single magnetic structure, in reality it can be more complicated.

We also note a couple of other interesting points -
Figure~\ref{fig-fitpar_20200406054800} shows that, unlike the other three elements, the abundance of Si does not quite reach the photospheric value during the episodic event. The same is true for other flares as well. 
The abundance values of Si near the flare peak (Table~\ref{table-I}, except the flares SOL2019-09-30T23:00, SOL2019-10-01T01:44, and SOL2020-04-04T00:52) are in good agreement with earlier large flare observation~\citep{veck_1981}.
It may be possible that at the chromospheric height where the evaporating plasma originates, the abundance of Si is already greater than that of the photosphere, indicating the fractionation of Si at lower chromospheric height. As a result, the evaporated plasma in the flaring loop exhibits Si abundance between photospheric and the coronal values, whereas the other elements exhibit photospheric abundance.
 
\begin{figure}[h!]
\begin{center}
\plottwo{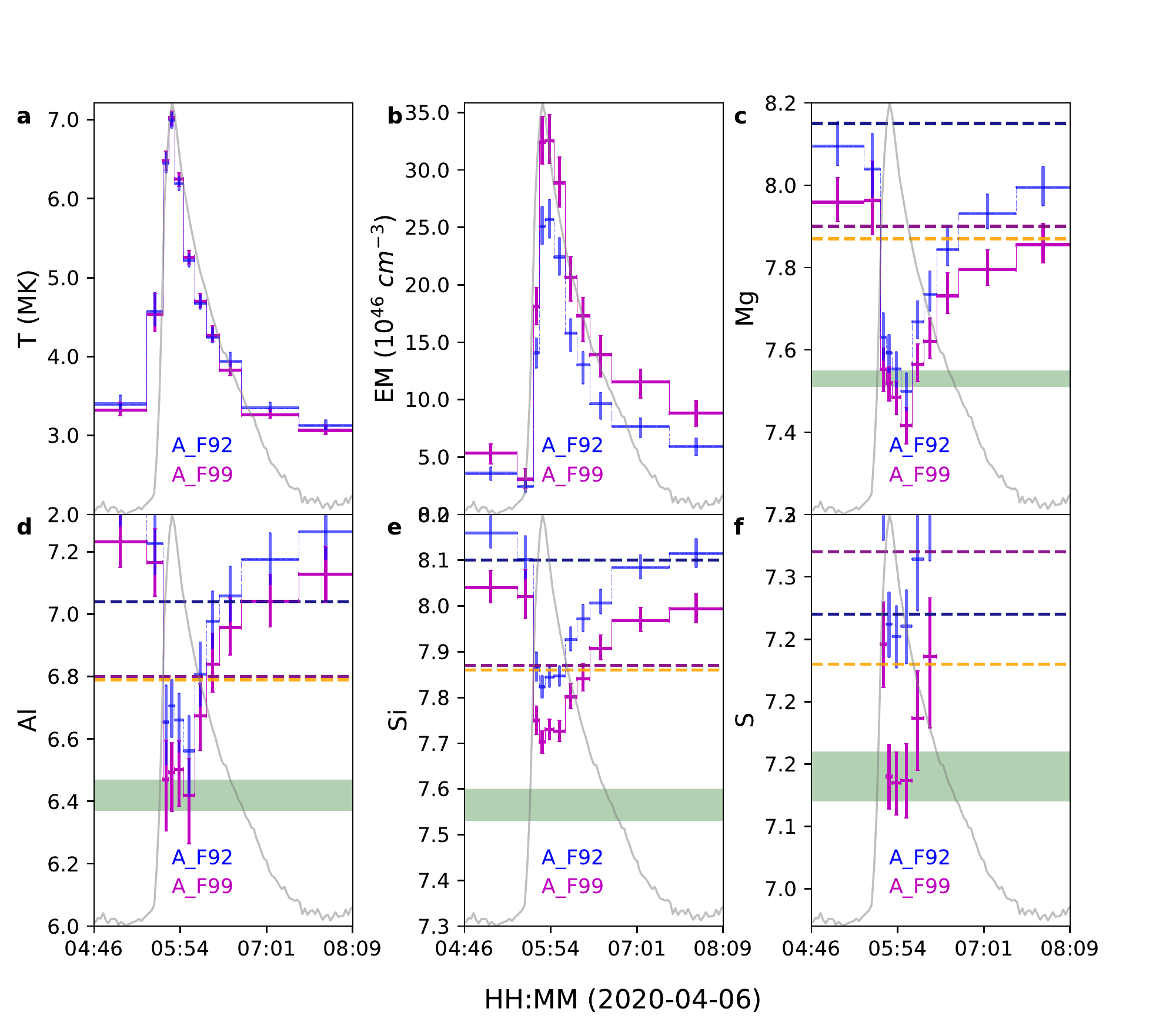}{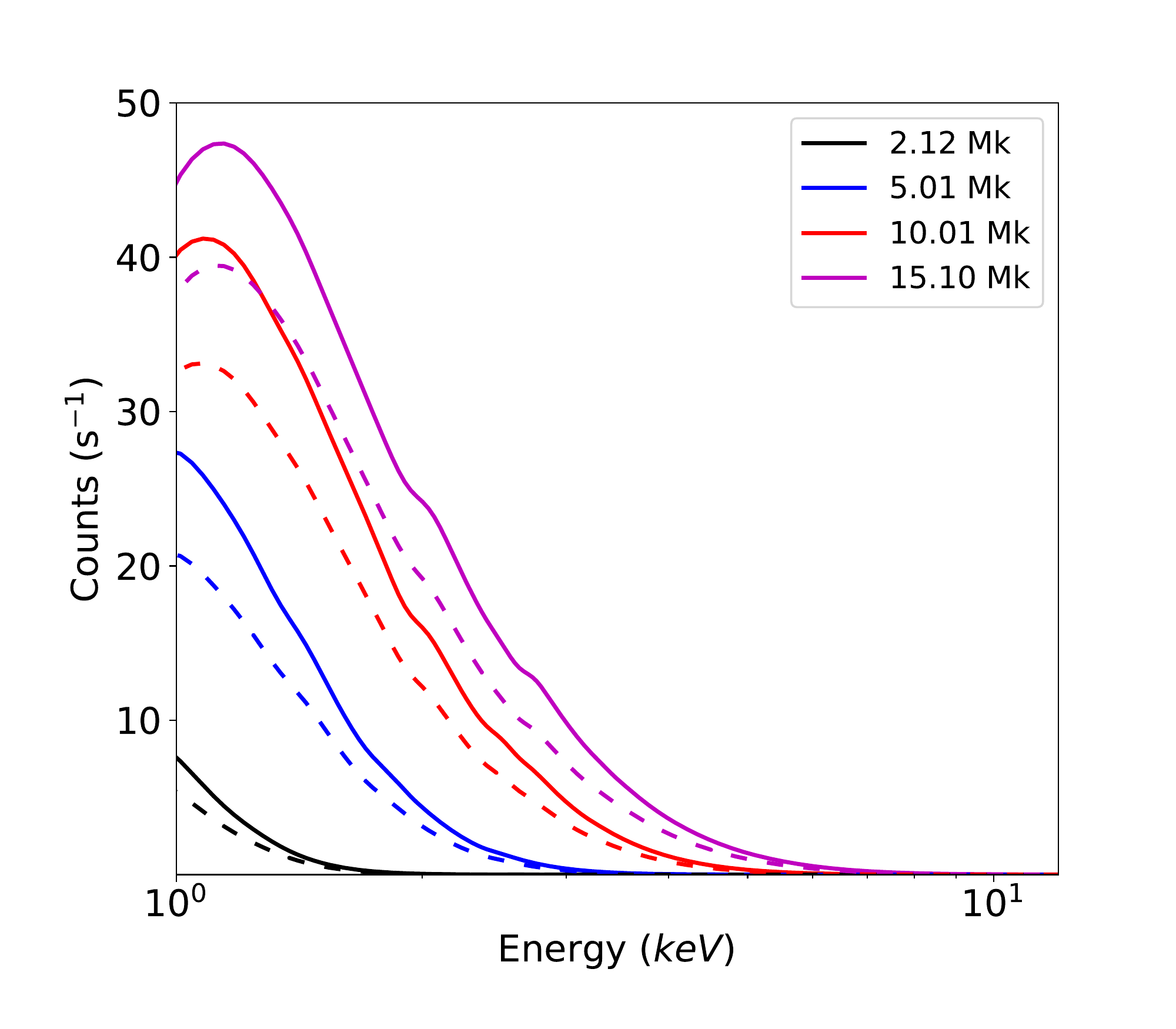}
\caption{The six panels on the left show the results of the time resolved X-ray spectral fitting using a single component model for the flare  SOL2020-04-06T05:48. 
The blue points with error bars show the fitted parameters by taking a fixed abundance parameter from A\_F92 and the magenta points with error bars show the same by using abundances values from A\_F99. 
The panel on the right represent  
the continuum spectrum at different temperature with a fixed emission measure of 2.0e47 cm$^{-3}$ by using the elemental abundances from A\_F92 (solid lines) and A\_F99 (dashed lines).}\label{fig-fitWithDiffAbund}
\end{center}
\end{figure}

It has been seen that the abundance of Mg during flare peaks depletes down to sub-photospheric values (Figures~\ref{fig-fitpar_20200406054800}c, \ref{fig-fitpar_AllMg}) on many occasions. Such abundance depletion may be mistaken as evidence of the inverse FIP effect. However, we find that this could be due to the effect of input abundances of unresolved lines in the continuum spectra. 
When we change the input abundances from A\_F99 (CHIANTI abundance dataset: ``sun\_coronal\_fludra\_1999\_ext.abund") to A\_F92 (CHIANTI abundane dataset: ``sun\_coronal\_feldman\_1992\_ext.abund") the abundance of Mg remains within the photospheric range.
For a similar reason, the measured abundances of flares SOL2019-09-30T23:00 (Panel a of Fgure~\ref{fig-fitpar_AllMg},\ref{fig-fitpar_AllSi}), SOL2019-10-01T01:44 (Panel b of Fgure~\ref{fig-fitpar_AllMg},\ref{fig-fitpar_AllSi}), and SOL2020-04-04T00:52 (Panel d of Fgure~\ref{fig-fitpar_AllMg},\ref{fig-fitpar_AllSi}), record lower values than the other flares.
To check further how input abundances affect our derived parameters, we compare our single component fit results (Figure~\ref{fig-fitWithDiffAbund}) of a typical B-class flare SOL2020-04-06T05:48 for two sets of input abundances; namely,
A\_F92, and A\_F99.
The temperature values (Figure~\ref{fig-fitWithDiffAbund}(a)) for both the input abundances remain almost unaltered. However, the rest of the parameters maintain a nearly constant offset. This is due to elements having a higher presence in the solar atmosphere, like O, C, etc., besides H and He. They play a crucial role in determining the soft X-ray continuum. When changing the abundances, the overall continuum gets changed. 
Figure~\ref{fig-fitWithDiffAbund}(b) shows the soft X-ray continuum at different temperatures. It also shows how the continuum shifts when the abundance list is changed from 
A\_F92, and A\_F99.
Since the continuum gets changed, the derived parameters resulting from the continuum are also affected. Though the exact values of the absolute abundances depend on the chosen input abundance list, the abundance evolutionary trend during flares and the coronal abundance recovery time scale remain independent of the abundance list.

Having analyzed several B-class flares and identified their spectroscopic behaviour over time, we can conclude that these flares also follow the standard flare model. In a manner similar to their larger, more energetic counterparts, they also show near photospheric abundances during their evolution. During the post flare phases, the quick recovery of the coronal abundances on a time scale of
minutes can be successfully explained through the Ponderomotive force model. To study this issue further, a careful analysis of multi-wavelength imaging and spectroscopic observations is necessary. We have carried out a campaign of multi-wavelength observations (with Hinode/EIS and XRT) and the results will be discussed in follow-up papers.

 \begin{deluxetable}{c |c |c |c c c c c c}
 \tablecaption{Plasma parameters obtained from the spectral analysis.
 $^a$Flare ID correspond to the time at the peak of the flare in the format SOLyyyy-mm-ddThh:mm.
 $^b$Active region number defined by NOAA for the associated flares. 
 $^c$Model used for the spectral fitting. $1T$ indicates the single component isothermal model. $2T$ indicates the two component model, where one component represent the flaring plasma emission and other component represent the non-flaring (NF) plasma emission, whose fitted parameters are given in the third row of each flare group. 
 $^d$Maximum temperature for the flares and the associated time bin in UT.
 $^e$Maximum emission measure and the associated time bin in UT.
 $^f$Elemental abundances when the emission measure is peaking.
 Errors represent the 1$\sigma$ limits associated with the parameters.}
\label{table-I}
\tablehead{Flare ID$^a$  &NOAA$^b$ &Model$^c$ & max.$^{d}$ T   & max.$^{e}$ EM  & Mg$^{f}$ & Al$^{f}$ & Si$^{f}$ & S$^{f}$ \\
&AR & & $\times 10 ^{6}$K   &  $\times 10 ^{46}$cm$^{-3}$ &  & &  & \\
& & & (Time bin)  &  (Time bin) &  & &  & 
}
\startdata
SOL2019-09-30T23:00	&	$	12749$&	$1T$	&	$	6.78^{+0.10}_{-0.12}$&	$	19.4^{+1.56}_{-1.22}$&	$	7.42^{+0.04}_{-0.05}$&	$	6.49^{+0.09}_{-0.12}$&	$	7.55^{+0.02}_{-0.02}$&	$	6.93^{+0.03}_{-0.03}$\\
&	&	&	($	22:54$-$23:03$)&	($	22:54$-$23:03$)&	&	&	&	\\
&	&	$2T$	&	$-$	&	$-$	&	$-$	&	$-$	&	$-$	&	$-$	\\
&	&	NF	&	$-$	&	$-$	&	$-$	&	$-$	&	$-$	&	$-$	\\
\cline{1-9}SOL2019-10-01T01:44	&	$	12749$&	$1T$	&	$	7.38^{+0.13}_{-0.17}$&	$	16.2^{+2.04}_{-1.32}$&	$	7.42^{+0.05}_{-0.07}$&	$	6.49^{+0.10}_{-0.17}$&	$	7.62^{+0.02}_{-0.03}$&	$	7.02^{+0.03}_{-0.04}$\\
&	&	&	($	01:42$-$01:46$)&	($	01:46$-$01:51$)&	&	&	&	\\
&	&	$2T$	&	$-$	&	$-$	&	$-$	&	$-$	&	$-$	&	$-$	\\
&	&	NF	&	$-$	&	$-$	&	$-$	&	$-$	&	$-$	&	$-$	\\
\cline{1-9}SOL2020-03-11T05:56	&	$	12758$&	$1T$	&	$	6.54^{+0.07}_{-0.10}$&	$	36.8^{+2.46}_{-1.97}$&	$	7.49^{+0.03}_{-0.04}$&	$	6.50^{+0.08}_{-0.11}$&	$	7.72^{+0.02}_{-0.02}$&	$	7.09^{+0.02}_{-0.02}$\\
&	&	&	($	05:44$-$05:52$)&	($	05:52$-$05:59$)&	&	&	&	\\
&	&	$2T$	&	$	6.58^{+0.08}_{-0.10}$&	$	36.4^{+2.37}_{-2.04}$&	$	7.47^{+0.03}_{-0.04}$&	$	6.48^{+0.08}_{-0.11}$&	$	7.71^{+0.02}_{-0.02}$&	$	7.09^{+0.02}_{-0.02}$\\
&	&	&	($	05:44$-$05:52$)&	($	05:52$-$05:59$)&	&	&	&	\\
&	&	NF	&	$	2.92^{+0.06}_{-0.04}$&	$	5.20^{+0.65}_{-0.77}$&	$	7.91^{+0.05}_{-0.04}$&	$	7.20^{+0.08}_{-0.08}$&	$	7.99^{+0.03}_{-0.03}$&	$-$	\\
\cline{1-9}SOL2020-04-04T00:52	&	$	12759$&	$1T$	&	$	9.06^{+0.08}_{-0.14}$&	$	42.3^{+3.20}_{-2.54}$&	$	7.31^{+0.04}_{-0.05}$&	$	6.41^{+0.10}_{-0.14}$&	$	7.55^{+0.02}_{-0.02}$&	$	7.02^{+0.02}_{-0.02}$\\
&	&	&	($	00:50$-$00:53$)&	($	00:53$-$00:57$)&	&	&	&	\\
&	&	$2T$	&	$	9.06^{+0.08}_{-0.16}$&	$	41.3^{+3.07}_{-2.65}$&	$	7.26^{+0.05}_{-0.05}$&	$	6.38^{+0.11}_{-0.15}$&	$	7.54^{+0.02}_{-0.02}$&	$	7.02^{+0.02}_{-0.02}$\\
&	&	&	($	00:50$-$00:53$)&	($	00:53$-$00:57$)&	&	&	&	\\
&	&	NF	&	$	2.99^{+0.04}_{-0.03}$&	$	8.41^{+0.67}_{-0.94}$&	$	7.83^{+0.04}_{-0.03}$&	$	7.11^{+0.07}_{-0.06}$&	$	7.98^{+0.02}_{-0.02}$&	$-$	\\
\cline{1-9}SOL2020-04-06T05:48	&	$	12759$&	$1T$	&	$	7.02^{+0.09}_{-0.10}$&	$	32.5^{+2.31}_{-1.98}$&	$	7.48^{+0.03}_{-0.04}$&	$	6.50^{+0.09}_{-0.11}$&	$	7.73^{+0.02}_{-0.02}$&	$	7.13^{+0.02}_{-0.02}$\\
&	&	&	($	05:45$-$05:50$)&	($	05:50$-$05:56$)&	&	&	&	\\
&	&	$2T$	&	$	7.08^{+0.09}_{-0.10}$&	$	31.5^{+2.17}_{-2.05}$&	$	7.44^{+0.04}_{-0.04}$&	$	6.45^{+0.10}_{-0.12}$&	$	7.72^{+0.02}_{-0.02}$&	$	7.13^{+0.02}_{-0.02}$\\
&	&	&	($	05:45$-$05:50$)&	($	05:50$-$05:56$)&	&	&	&	\\
&	&	NF	&	$	3.13^{+0.04}_{-0.03}$&	$	7.21^{+0.55}_{-0.66}$&	$	7.84^{+0.03}_{-0.03}$&	$	7.12^{+0.06}_{-0.05}$&	$	7.97^{+0.02}_{-0.01}$&	$-$	\\
\cline{1-9}SOL2020-04-06T08:32	&	$	12759$&	$1T$	&	$	7.40^{+0.10}_{-0.15}$&	$	23.0^{+2.88}_{-1.80}$&	$	7.51^{+0.04}_{-0.06}$&	$	6.37^{+0.13}_{-0.24}$&	$	7.76^{+0.02}_{-0.03}$&	$	7.08^{+0.03}_{-0.04}$\\
&	&	&	($	08:30$-$08:34$)&	($	08:34$-$08:39$)&	&	&	&	\\
&	&	$2T$	&	$	7.66^{+0.15}_{-0.18}$&	$	21.5^{+2.69}_{-1.83}$&	$	7.45^{+0.05}_{-0.06}$&	$	6.26^{+0.17}_{-0.35}$&	$	7.75^{+0.02}_{-0.03}$&	$	7.07^{+0.03}_{-0.04}$\\
&	&	&	($	08:26$-$08:30$)&	($	08:34$-$08:39$)&	&	&	&	\\
&	&	NF	&	$	3.13^{+0.04}_{-0.03}$&	$	7.21^{+0.55}_{-0.66}$&	$	7.84^{+0.03}_{-0.03}$&	$	7.12^{+0.06}_{-0.05}$&	$	7.97^{+0.02}_{-0.01}$&	$-$	\\
\cline{1-9}SOL2020-04-07T01:58	&	$	12759$&	$1T$	&	$	8.08^{+0.05}_{-0.11}$&	$	47.6^{+3.18}_{-2.57}$&	$	7.51^{+0.03}_{-0.04}$&	$	6.48^{+0.09}_{-0.12}$&	$	7.71^{+0.02}_{-0.02}$&	$	7.08^{+0.02}_{-0.02}$\\
&	&	&	($	01:56$-$02:01$)&	($	02:01$-$02:05$)&	&	&	&	\\
&	&	$2T$	&	$	8.11^{+0.05}_{-0.11}$&	$	47.0^{+3.07}_{-2.64}$&	$	7.49^{+0.03}_{-0.04}$&	$	6.46^{+0.09}_{-0.12}$&	$	7.71^{+0.02}_{-0.02}$&	$	7.08^{+0.02}_{-0.02}$\\
&	&	&	($	01:56$-$02:01$)&	($	02:01$-$02:05$)&	&	&	&	\\
&	&	NF	&	$	2.89^{+0.04}_{-0.03}$&	$	7.68^{+0.65}_{-0.80}$&	$	7.78^{+0.04}_{-0.03}$&	$	7.01^{+0.08}_{-0.08}$&	$	7.92^{+0.02}_{-0.02}$&	$-$	\\
\cline{1-9}SOL2020-04-09T09:32	&	$	12759$&	$1T$	&	$	6.44^{+0.09}_{-0.16}$&	$	28.8^{+1.84}_{-1.78}$&	$	7.46^{+0.03}_{-0.03}$&	$	6.50^{+0.09}_{-0.10}$&	$	7.71^{+0.02}_{-0.02}$&	$	7.12^{+0.02}_{-0.02}$\\
&	&	&	($	09:08$-$09:13$)&	($	09:36$-$09:47$)&	&	&	&	\\
&	&	$2T$	&	$	6.52^{+0.10}_{-0.16}$&	$	28.1^{+1.79}_{-1.76}$&	$	7.42^{+0.03}_{-0.03}$&	$	6.46^{+0.09}_{-0.11}$&	$	7.71^{+0.02}_{-0.02}$&	$	7.12^{+0.02}_{-0.02}$\\
&	&	&	($	09:08$-$09:13$)&	($	09:36$-$09:47$)&	&	&	&	\\
&	&	NF	&	$	2.85^{+0.04}_{-0.03}$&	$	6.71^{+0.67}_{-0.76}$&	$	7.83^{+0.04}_{-0.03}$&	$	7.11^{+0.08}_{-0.07}$&	$	7.91^{+0.02}_{-0.02}$&	$-$	\\
\cline{1-9}SOL2020-04-30T12:16	&	$	12762$&	$1T$	&	$	7.39^{+0.14}_{-0.18}$&	$	15.5^{+2.14}_{-1.88}$&	$	7.46^{+0.06}_{-0.06}$&	$	6.44^{+0.15}_{-0.22}$&	$	7.68^{+0.03}_{-0.03}$&	$	7.13^{+0.05}_{-0.05}$\\
&	&	&	($	12:15$-$12:18$)&	($	12:18$-$12:22$)&	&	&	&	\\
&	&	$2T$	&	$	7.69^{+0.17}_{-0.18}$&	$	12.4^{+1.40}_{-1.19}$&	$	7.48^{+0.06}_{-0.07}$&	$	6.64^{+0.11}_{-0.15}$&	$	7.67^{+0.03}_{-0.03}$&	$	7.12^{+0.04}_{-0.04}$\\
&	&	&	($	12:15$-$12:18$)&	($	12:15$-$12:18$)&	&	&	&	\\
&	&	NF	&	$	3.15^{+0.05}_{-0.03}$&	$	8.98^{+0.70}_{-1.01}$&	$	7.84^{+0.04}_{-0.03}$&	$	7.15^{+0.07}_{-0.05}$&	$	7.94^{+0.02}_{-0.02}$&	$-$	\\
\cline{1-9}
\enddata
\end{deluxetable}

\section{Summary}\label{discussion}

In this article, we presented the temporal evolution of elemental abundances during solar flares of GOES B1-B4 class, the weakest events for which such studies have been possible so far, using observations with XSM on board Chandrayaan-2.
 This study was possible due to the extremely quiet solar conditions currently prevailing  
along with the 
availability of appropriate instrumentation. 
 By modelling the soft X-ray spectra obtained with the
 XSM during different phases of these flares, measurements of temperature, emission measure, and abundances of Mg, Al,
 Si, and S were obtained. 
 We have shown that the abundances of these elements are nearly photospheric
 during the peak phase, compared to their 3-4 times higher pre-flare values. 
 Beyond the flare peak, the abundances are seen to get enriched again
 and they recover back to the coronal values at the end of the flare. This suggests that during the flares, the coronal loops
 are quickly filled with plasma originating from the lower parts of the solar atmosphere without having sufficient time for the
 usually observed fractionation in non-flaring loops to take place.
 Our observation of quick recovery to the coronal values show that any process giving rise to such fraction must be occurring on a time scale of few tens of minutes.

\acknowledgments{
We are thankful to an anonymous referee for providing us with very useful feedback.
XSM was designed and developed by the Physical Research Laboratory (PRL), Ahmedabad
with support from the  Space Application Centre (SAC), Ahmedabad,
the U. R. Rao Satellite Centre (URSC), Bengaluru, and the Laboratory for Electro-Optics
Systems (LEOS), Bengaluru.
We thank various facilities and the technical teams of all the above centers 
and the Chandrayaan-2 project, mission operations, and ground segment teams for 
their support. The Chandrayaan-2 mission is funded and managed by the Indian Space Research Organisation (ISRO).
Research at PRL is supported by the Department of Space, Govt. of India. 
The collaboration with GDZ, HEM, and UMK at DAMTP, Cambridge Uni. is facilitated through the Royal Society Grant No. IES{\textbackslash}R2{\textbackslash}170199.
GDZ and HEM acknowledge support from STFC (UK) via the consolidated grants 
to the atomic astrophysics group (AAG) at DAMTP, University of Cambridge
(ST/P000665/1 and ST/T000481/1).}

\vspace{5mm}
\facilities{Chandrayaan-2 (XSM)}
\software{XSMDAS~\citep{xsm_data_processing_2020}, XSPEC~\citep{ref-xspec}, Python, Matplotlib}

\appendix 
\renewcommand\thefigure{\thesection.\arabic{figure}}
\setcounter{figure}{0}
\section{An XSPEC model for ionized plasma emission using CHIANTI}\label{sec-model}

In this work, we model the observed soft X-ray spectrum of solar flares with the line and 
continuum emission from an isothermal plasma characterized by a temperature, emission measure,
and abundances of various elements, computed using the CHIANTI atomic 
database~\citep{chiantiV10_Zanna2020}.
As such a model is not readily available in the X-ray spectral fitting package XSPEC~\citep{ref-xspec} used for the analysis, we have incorporated it as a local model named \emph{chisoth}.

While it is possible to include this as a model in PyXspec, the pythonic version of XSPEC, by using the functions available in CHIANTIPy, execution time for each model calculation is prohibitively large. Another possibility is to use the option of table model in XSPEC, where the model spectra over a multi-dimensional grid of parameters are stored in a file and spectrum for any required set of parameters is obtained by interpolation. However, this is also not a viable option due to the requirement of abundances of elements being independent parameters resulting in very large file sizes. Thus, we adopt a different approach and store a library of spectra only over a grid of temperatures, but for each individual element. Then, the total model spectrum for a required temperature and abundances is obtained by weighted addition of interpolated spectra of each elements, as detailed below. The pre-computed element-wise spectra are recorded in a FITS file and the model is implemented in C++, which can be compiled and loaded into XSPEC following the standard procedure\footnote{\url{https://heasarc.gsfc.nasa.gov/xanadu/xspec/manual/XSappendixLocal.html}}.

In order to generate the spectral library, line and continuum spectra for each element (Z=1 to 30) at unity abundance were computed over a logarithmic grid ($\delta$log(T) = 0.004 K)
of temperatures from 0.3 MK to 50 MK
using CHIANTI routines. Line spectra were obtained by using $ch\_synthetic.pro$, while the continuum spectra were obtained using individual routines for different emission processes, and all were added together to obtain the total spectra for each element at different temperatures, which are recorded in a FITS file. 

Within the XSPEC model, the spectrum for each element at any temperature $T$ within the range is obtained by linear interpolation between the spectra at the nearest two grid points available in the spectral library. Then the total spectrum $I(T,E)$ is obtained by weighted addition of spectra of individual elements as:
\begin{equation}\label{spe_eq}
I(T,E)=EM * [I_{\rm {H}}(T,E) + I_{\rm {He}}(T,E) * N_{\rm {He}} +....+ I_{\rm Fe}(T,E) * N_{\rm {Fe}} +.....+I_{\rm {Zn}}(T,E) * N_{\rm {Zn}}]
\end{equation}
where $EM$ is the volume emission measure, $I_{\rm {X}}(T,E)$ is spectrum for element X at unity abundance  and emission measure and $N_{\rm{X}}$ is the abundance of element X relative to that of H, which is related to the usually used logarithmic value $A_{\rm{X}}$ as 
\begin{equation}\label{abund_eq}
A_{\rm{X}} = 12 + log_{10}(N_{\rm{X}})
\end{equation}
The model takes logarithm of temperature (logT), abundances ($A_{\rm{X}}$) of the elements with Z=2 to Z=30, and volume emission measure as input parameters.
It may be noted that the the volume emission measure is implemented as an overall normalization factor, which is in units of $10^{46} {~\rm{cm}}^{-3}$. Using the features in XSPEC, it is possible to freeze specific parameters such as abundances of certain elements to fixed values while keeping others free during spectral fitting. It is also possible to fit the spectrum with a sum of multiple isothermal emission models as used in this work. 

\begin{figure}[ht!]
\centering
\includegraphics[width=0.8\linewidth]{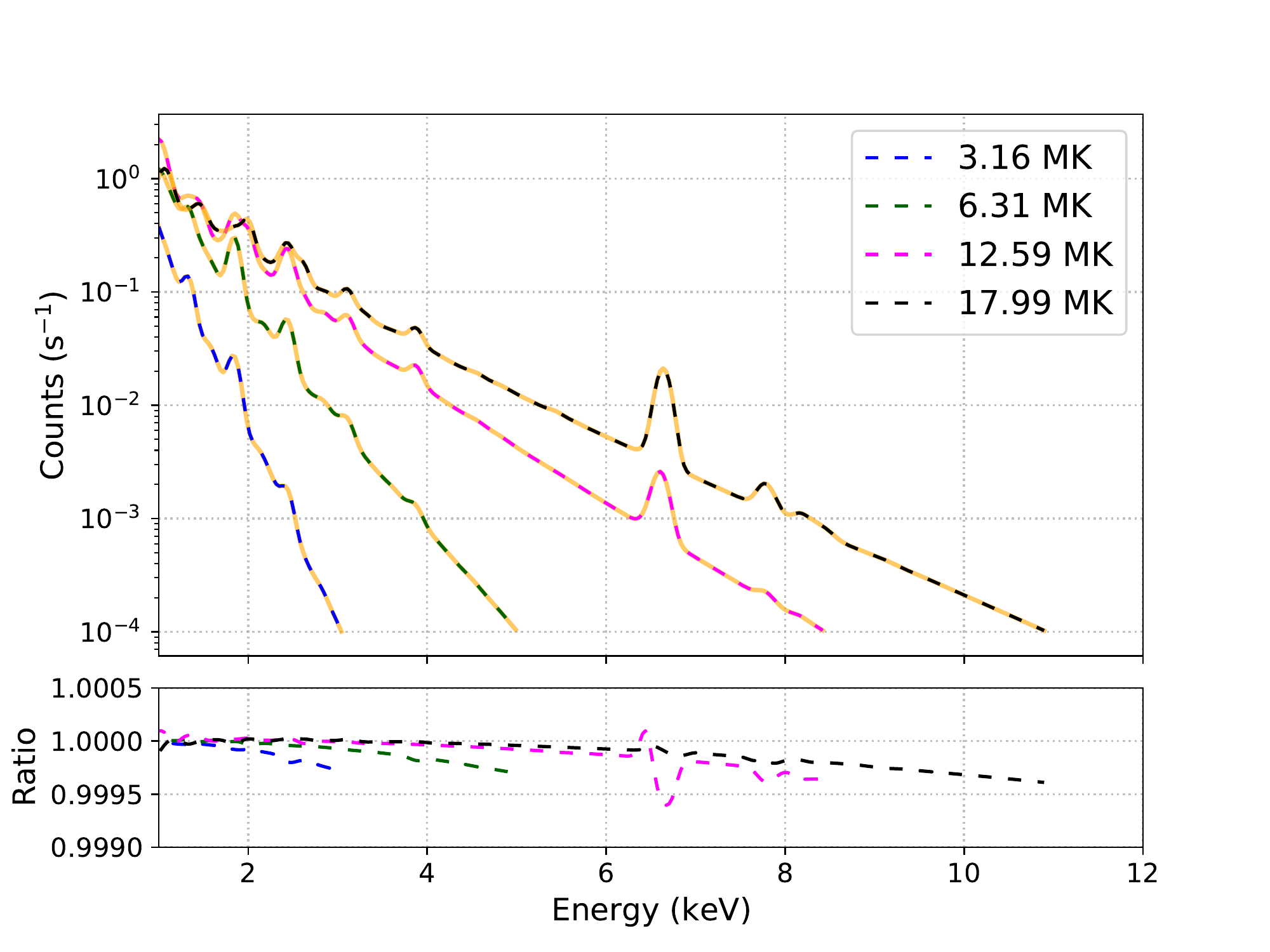}
\caption{Comparison of interpolated spectrum from XSPEC model with that from direct calculations with CHIANTI. The upper panel shows the generated spectra for four different temperatures using CHIANTI (orange dashed line), over-plotted with the spectrum obtained from XSPEC 
(dashed lines of blue, green, magenta, and black colors) and the lower panel shows the ratio between the two. 
}
\label{fig-mo_valid}
\end{figure}

In order to verify the adequacy of the grid of temperatures used in the model, we carried out 
comparison of the spectra obtained from the XSPEC model at non-grid temperature values with 
that obtained directly from CHIANTI. Figure~\ref{fig-mo_valid} shows such comparison at four 
representative temperatures. The upper panel shows the spectra obtained from the XSPEC model 
and CHIANTI, convolved with the XSM response and the lower panel shows the ratio between the 
two. It can be seen from the figure that the differences are less than 0.1\%, demonstrating 
the adequacy of the XSPEC model for analysis of broad-band soft X-ray spectra of the Sun. 

\newpage
\bibliography{myref}   
\bibliographystyle{apalike} 

\end{document}